\documentclass[12pt]{article}
\pdfoutput=1

\usepackage{putex}
\usepackage{graphicx}
\usepackage{caption}
\usepackage{amsmath}
\usepackage{array}
\usepackage{subcaption}
\usepackage{epstopdf}
\usepackage{enumerate}
\usepackage{cite}
\usepackage{youngtab}
\usepackage{tensor}
\usepackage{slashed}
\usepackage[aligntableaux=center]{ytableau}
\usepackage[utf8]{inputenc}
\usepackage{rotating}
\usepackage{bigfoot}
\usepackage[
      colorlinks=true,
      linkcolor=blue,
      urlcolor=blue,
      filecolor=black,
      citecolor=red,
      ]{hyperref}

\newcommand{\abs}[1]{\left\lvert #1 \right\rvert}

\newcommand {\be} {\begin {equation}}
\newcommand {\ee} {\end {equation}}

\newcommand {\bes} {\begin {equation*}}
\newcommand {\ees} {\end {equation*}}

\newcommand{\es}[2] {\begin{equation} \label{#1} \begin{split} #2 \end{split} \end{equation}}

\newcommand{\Z}{\mathbb{Z}}

\newcommand{\R}{\mathbb{R}}

\newcommand{\cA}{{\mathcal A}}
\newcommand{\cB}{{\mathcal B}}

\newcommand{\cF}{{\mathcal F}}
\newcommand{\cG}{{\mathcal G}}

\newcommand{\cN}{{\mathcal N}}
\newcommand{\cO}{{\mathcal O}}
\newcommand{\cP}{{\mathcal P}}

\newcommand{\cS}{{\mathcal S}}
\newcommand{\cT}{{\mathcal T}}

\newcommand{\cM}{{\mathcal M}}

\newcommand{\beq}{\begin{equation}}
\newcommand{\eeq}{\end{equation}}

\def\ie{\begin{equation}\begin{aligned}}
\def\fe{\end{aligned}\end{equation}}

\numberwithin{equation}{section}

\def\<{\langle}
\def\>{\rangle}

\begin{document}

\preprint{PUPT-2616}

\institution{Exile}{Department of Particle Physics and Astrophysics, Weizmann Institute of Science, Rehovot, Israel}
\institution{PU}{Joseph Henry Laboratories, Princeton University, Princeton, NJ 08544, USA}

\title{Far Beyond the Planar Limit in Strongly-Coupled $\mathcal{N}=4$ SYM}

\authors{Shai M.~Chester\worksat{\Exile} and Silviu S.~Pufu\worksat{\PU}}

\abstract{
When the $SU(N)$ ${\cal N} = 4$ super-Yang-Mills (SYM) theory with complexified gauge coupling $\tau$ is placed on a round four-sphere and deformed by an ${\cal N} = 2$-preserving mass parameter $m$, its free energy $F(m, \tau, \bar \tau)$ can be computed exactly using supersymmetric localization.  In this work, we derive a new exact relation between the fourth derivative $\partial_m^4 F(m, \tau, \bar \tau) \big|_{m=0}$ of the sphere free energy and the integrated stress-tensor multiplet four-point function in the $\mathcal{N}=4$ SYM theory.   We then apply this exact relation, along with various other constraints derived in previous work (coming from analytic bootstrap, the mixed derivative $\partial_\tau \partial_{\bar \tau} \partial_m^2 F(m, \tau, \bar \tau) \big|_{m=0}$, and type IIB superstring theory scattering amplitudes) to determine various perturbative terms in the large $N$ and large 't Hooft coupling $\lambda$ expansion of the ${\cal N} = 4$ SYM correlator at separated points.  In particular, we determine the leading large-$\lambda$ term in the ${\cal N} = 4$ SYM correlation function at order $1/N^8$.  This is three orders beyond the planar limit.
}
\date{March 2018}

\maketitle

\tableofcontents

\section{Introduction}
\label{intro}

The four-point functions of stress-tensor multiplet operators in the $SU(N)$ $\mathcal{N}=4$ super-Yang-Mills (SYM) theory have received a significant amount of attention over the past twenty or so years.   For small Yang-Mills coupling $g_\text{YM}$, these correlators can be computed perturbatively for any $N$ using standard Feynman diagrams (see \cite{GonzalezRey:1998tk,Eden:2000mv,Bianchi:2000hn,Drummond:2013nda} for expressions up to three loops). At large $N$ and large 't Hooft coupling $\lambda \equiv g_\text{YM}^2N$, they can in principle be computed using Witten diagrams in an expansion around classical type IIB supergravity on $AdS_5\times S^5$~\cite{Maldacena:1997re,Gubser:1998bc,Witten:1998qj}. In this limit, $1/\lambda$ corrections correspond to higher derivative terms in the effective action that correct the two-derivative supergravity action, while $1/N$ corrections correspond, roughly, to loop diagrams.\footnote{More precisely, at subleading orders in $1/N$ there are both contributions from loop diagrams and from tree-level diagrams.  Some of the tree-level contributions can be separated out from the loop contributions because they have a different scaling in $1/\lambda$.  We will provide examples in the next section.} At leading order in $1/\lambda$ and $1/N$, i.e.~in tree-level supergravity, the connected stress-tensor multiplet correlators are known from explicit Witten diagram computations \cite{Eden:2000bk,Arutyunov:2002fh,Arutyunov:2003ae,Berdichevsky:2007xd,Uruchurtu:2008kp,Uruchurtu:2011wh,Arutyunov:2017dti,Arutyunov:2018tvn}, but this approach becomes difficult to pursue at loop level or for higher derivative corrections to supergravity, partly because loop computations in AdS are complicated, and partly because even the interaction vertices corresponding to the first higher-derivative correction to supergravity are not fully known (see however \cite{deHaro:2002vk,Policastro:2006vt,Paulos:2008tn,Liu:2013dna} for partial results). Recently, this obstacle has been overcome using a combination of techniques:  analytic bootstrap \cite{Heemskerk:2009pn,Rastelli:2017udc},\footnote{See \cite{Caron-Huot:2018kta,Goncalves:2019znr,Rastelli:2017ymc,Rastelli:2019gtj,Zhou:2017zaw,Zhou:2018ofp,Alday:2014tsa,Chester:2018lbz,Chester:2018aca,Chester:2018dga,Binder:2018yvd,Binder:2019mpb,Aprile:2018efk,Giusto:2018ovt,Giusto:2019pxc,Alday:2017gde,Alday:2019qrf} for other applications of these methods to holographic correlators in various dimensions.} supersymmetric localization \cite{Pestun:2007rz,Binder:2019jwn,Chester:2019pvm,Chester:2019jas}, the flat space limit \cite{Polchinski:1999ry,Susskind:1998vk,Giddings:1999jq,Fitzpatrick:2011hu,Penedones:2010ue,Fitzpatrick:2011jn,Goncalves:2014ffa}, and unitarity methods \cite{Aharony:2016dwx,Alday:2017xua,Aprile:2017bgs,Aprile:2017xsp,Aprile:2017qoy,Alday:2018pdi,Alday:2018kkw,Drummond:2019odu,Alday:2017vkk,Drummond:2019hel,Aprile:2019rep,Alday:2019nin}, which do not require detailed knowledge of the bulk action. 

In this work, we will derive a new relation between the integrated stress-tensor multiplet correlator and four mass derivatives 
 \es{calF1Def}{
   \cF_4(\tau, \bar \tau) \equiv \frac{\partial^4 F(m,\tau, \bar \tau)}{\partial m^4} \bigg\vert_{m=0}
 }
of the free energy $F(m, \tau, \bar \tau)$ of the ${\cal N} = 2^*$ theory placed on a round four-sphere.   (The ${\cal N} = 2^*$ theory is a mass deformation of the ${\cal N} = 4$ SYM theory that preserves ${\cal N} = 2$ supersymmetry.  It depends on the mass parameter $m$ as well as the complexified gauge coupling  $\tau\equiv \frac{\theta}{2\pi}+\frac{4\pi i}{g_\text{YM}^2}$ and its conjugate $\bar \tau$.)  The relation we derive is an extension of a similar relation between the stress-tensor multiplet correlator and the mixed fourth derivative 
 \es{calF2Def}{
   \cF_2(\tau, \bar \tau) \equiv \frac{\partial^4 F(m,\tau, \bar \tau)}{\partial \tau \partial \bar \tau \partial m^2} \bigg\vert_{m=0}
 }
that was previously studied in \cite{Binder:2019jwn}. Since $F(m,\tau, \bar \tau)$ can be computed using supersymmetric localization \cite{Pestun:2007rz}, both the relation derived here and that of \cite{Binder:2019jwn}  impose non-perturbative constraints on the stress-tensor multiplet correlator for any $N$ and $(\tau, \bar \tau)$.  As an application, we will use these constraints to derive new terms in the perturbative $1/N$ and $1/\lambda$ expansion of the stress tensor correlator, as we will describe shortly.  

In more detail, the stress tensor multiplet of the ${\cal N} = 4$ SYM theory contains 42 real scalar operators:  $20$ of them, which we collectively denote by $S$, have scaling dimension~$2$, transform as the ${\bf 20}'$ of the $SU(4)_R$ R-symmetry, and in the Lagrangian description are single trace scalar bilinears;  another $20$ operators, grouped in the complex combinations $P$ and $\overline{P}$, have scaling dimension $3$, transform in the ${\bf 10}$ and $\overline{\bf 10}$, respectively, of $SU(4)_R$, and in the Lagrangian description are single trace fermion bilinears;  and lastly $2$ operators have dimension $4$ that will not be important in this paper.  Four-point correlation functions of all these operators, as well as of other operators belonging to the ${\cal N} = 4$ stress tensor multiplet, are related to one another by Ward identities, and can all be expressed in terms of a single function $\cT(U, V)$ of the conformally-invariant cross-ratios $U$ and $V$ \cite{Dolan:2001tt}.  

Coming back to the derivatives $\cF_4(\tau, \bar \tau)$ and $\cF_2(\tau, \bar \tau)$ of the $S^4$ free energy, these quantities can be related to $\cT(U, V)$ because each derivative w.r.t.~$m$ corresponds to the insertion of a specific linear combination of $S$, $P$, and $\overline{P}$ integrated over the four-sphere, while a derivative w.r.t.~$\tau$ (or $\bar \tau$) corresponds \cite{Gerchkovitz:2016gxx,Gomis:2014woa,Gerchkovitz:2014gta} to an insertion of a specific component of $S$ at the north (or south) pole of the sphere.  Thus, $\cF_2(\tau, \bar \tau)$ can be written in terms of $\langle SSSS \rangle$ and $\langle SS P \overline{P} \rangle$, where two of the $S$ operators in the first correlator as well as the $P$ and $\overline{P}$ operators in the second correlator are integrated.  Writing these integrated correlators in terms of integrals of ${\cal T}(U, V)$ was achieved in \cite{Binder:2019jwn}, following a similar calculation in the 3d Aharony-Bergman-Jafferis-Maldacena (ABJM) theory \cite{Aharony:2008ug} described in \cite{Binder:2018yvd}.  In this paper, we will perform the same task for $\cF_4(\tau, \bar \tau)$, which can be written as a linear combination of $\langle SSSS\rangle$, $\langle SSP\overline P\rangle$, and $\langle P\overline PP\overline P\rangle$, where now all four operators are integrated over the sphere.   This calculation has two challenges.  The first is to write $\langle SS P \overline{P}\rangle$ and $\langle P\overline PP\overline P\rangle$ in terms of ${\cal T}(U, V)$ using the Ward identities, which we do following the component method also used in \cite{Dolan:2001tt,Binder:2019jwn}.\footnote{The $\langle SSP\overline P\rangle$ Ward identity was already derived in \cite{Binder:2019jwn}. The solution to the Ward identities relating $\langle SS P \overline{P}\rangle$ and $\langle P\overline PP\overline P\rangle$ to $\langle SSSS \rangle$ can in principle also be read off from \cite{Belitsky:2014zha}, where a super-space expression of the stress tensor multiplet correlators in terms of the function ${\cal T}(U, V)$ was given.}   The second challenge is to perform the integrals over the sphere, where, unlike in the case of $\cF_2(\tau, \bar \tau)$, one now encounters additional divergences that need to be regularized while preserving supersymmetry.

To use the relation between $\cF_4$ and ${\cal T}(U, V)$ in the holographic regime, we should derive an expansion of $\cF_4(\tau, \bar \tau)$ at large $N$ and large $\lambda$. Using Ref.~\cite{Pestun:2007rz} as a starting point, one can write down $\cF_4(\tau, \bar \tau)$ as an expectation value of an operator in the free Gaussian matrix model at $m=0$.   As in \cite{Chester:2019pvm}, this expectation value can then be computed to any order in $1/N$ at finite $\lambda$ using topological recursion \cite{Eynard:2004mh,Eynard:2008we}, and also at finite $N$ and $\lambda$ (if we ignore non-perturbative instantons in the Nekrasov partition function) using orthogonal polynomials \cite{mehta1981}. 

The expansion of $\cF_4$ in $1/N$ and $1/\lambda$, combined with various other constraints studied in previous work, can be used to fully determine the function $\cT(U, V)$ to higher orders in the double expansion in $1/N$ and $1/\lambda$ than was previously possible.  In particular, we determine that the Mellin transform\footnote{The precise definition of the Mellin transform is given in Eq.~\eqref{MellinDef} below.} \cite{Penedones:2010ue,Fitzpatrick:2011ia} of the function $\cT(U, V)$, which we denote by $\cM(s, t)$, takes the form
\es{MellinFinalIntro}{
   \cM(s, t) =&\frac1 c\left[\frac{8}{(s-2)(t-2)(u-2)}+\frac{120\zeta(3)}{\lambda^{\frac32}}   +\frac{630\zeta(5)}{\lambda^{\frac52}}\left[s^2+t^2+u^2-3\right]\right.\\
   &\left.\qquad+\frac{5040\zeta(3)^2}{\lambda^{3}}\left[stu-\frac14(s^2+t^2+u^2)-4\right]+O(\lambda^{-3})\right]\\
   &+\frac{1}{c^2}
   \left[\frac{5\sqrt{\lambda}}{8}+\mathcal{M}^{\text{SG}|\text{SG}}+\frac{15}{4} +O(\lambda^{-\frac32})\right]\\
   &+\frac{1}{c^3}
   \left[\frac{7\lambda^{\frac32}}{3072}\left[s^2+t^2+u^2 - 3  \right]+O(\lambda)\right]\\
    &+\frac{1}{c^4}
   \left[\frac{\lambda^{3}}{221184} \left[stu-\frac14(s^2+t^2+u^2)-4\right]+O(\lambda^{\frac52})\right]+O(c^{-5})\,,
}
where $u \equiv 4 - s - t$, and where $c = (N^2 - 1)/4$ is the $c$ anomaly coefficient, which is the natural expansion for holographic correlators since it is simply related to the effective 5d Newton's constant.  In string theory language, the terms at order $1/c^{g+1}$ correspond to genus~$\leq g$ string worldsheets, so the expansion \eqref{MellinFinalIntro} contains contributions up to genus three.  

The expression \eqref{MellinFinalIntro} was determined as follows:
 \begin{itemize}
  \item Crossing symmetry and the analytic structure of Witten diagrams in Mellin space \cite{Penedones:2010ue,Fitzpatrick:2011ia,Fitzpatrick:2011hu,Fitzpatrick:2011dm} determine the $s, t$ dependence of each term in the $1/c$ and $1/\lambda$ expansion in Eq.~\eqref{MellinFinalIntro} up to undetermined coefficients.  In particular, the polynomial terms in $s$, $t$, $u$ correspond to contact Witten diagrams, where for a polynomial of degree $n$, the interaction vertex is schematically of the form $D^{2n} R^4$;  the first term at order $1/c$ corresponds to the tree-level supergravity amplitude;  and the $\mathcal{M}^{\text{SG}|\text{SG}}$ term corresponds to the one-loop supergravity amplitude, which is a non-analytic in $s$, $t$, $u$, and was determined in \cite{Aprile:2017bgs,Alday:2017xua,Alday:2018kkw} using unitarity, up to an additive constant.
  \item The coefficient of the supergravity term is fixed by the requirement that, when expanding the full correlator in conformal blocks, there are no operators of dimension precisely two \cite{Rastelli:2017udc}.  
  \item At each order in the $1/c$ and $1/\lambda$ expansion, one can determine the coefficient of the leading term at large $s$, $t$, $u$ from knowledge of the flat space scattering amplitude in type IIB superstring theory.   This was originally done in \cite{Goncalves:2014ffa} to fully determine the term of order $c^{-1} \lambda^{-3/2}$ (i.e.~the genus zero $R^4$ term).
  \item At each order in $1/c$ and $1/\lambda$, one can determine two coefficients, namely one from $\cF_4$ and one from $\cF_2$, when these quantities are also expanded in $1/c$ and $1/\lambda$.  In particular, in \cite{Binder:2019jwn} the $\cF_2$ constraint was used to fully determine the term of order $c^{-1} \lambda^{-3/2}$ (genus zero $R^4$), and to also determine the remaining coefficient in the $c^{-1} \lambda^{-5/2}$ term (genus zero $D^4 R^4$) that remained undetermined after using the flat space limit.  In \cite{Chester:2019pvm}, the quantity $\cF_2$ was computed to any order in $1/N$ and $1/\lambda$, and used to also fix the $c^{-2} \lambda^{1/2}$, $c^{-3} \lambda^{3/2}$, and $c^{-2} \lambda^0$ terms.  The rest of the coefficients in \eqref{MellinFinalIntro} are determined in this paper.
  \end{itemize}

Note that the coefficients corresponding to $R^4$ and $D^4 R^4$ can be fully fixed using only the two supersymmetric localization constraints, and they do agree, in the flat space limit, with the scattering amplitude in type IIB superstring theory.  They appear at genus zero and genus one in the case of $R^4$, and at genus zero and genus two for $D^4 R^4$.   The match between supersymmetric localization and type IIB scattering amplitudes represents a non-trivial precision test of AdS/CFT at these orders.\footnote{The $R^4$ and $D^4R^4$ coefficients were also fixed in in \cite{Binder:2018yvd} for the ABJM holographic correlator, which is dual to M-theory on $AdS_4\times S^7$, using similar localization constraints. In that case, however, while the $R^4$ coefficient is non-zero, the $D^4R^4$ coefficient vanishes.}  The terms of order $1/c^4$ in \eqref{MellinFinalIntro} were obtained by combining the supersymmetric localization and flat space limit constraints, and they represent, to our knowledge, the first known contributions to a holographic correlator at genus three.  
  
  The rest of this paper is organized as follows.  In Section~\ref{4point}, we discuss the stress tensor multiplet four-point function in the strong coupling limit, and fix the higher order in $1/N$ and $1/\lambda$ terms using the flat space limit, the old $\cF_2$ constraint, and the new $\cF_4$ constraint. In Section~\ref{LocCon}, we derive this new integrated constraint. In the Appendices we include many details of the calculation, including the localization calculation of $\cF_4$ from topological recursion or orthogonal polynomials.  We end with a discussion of our results and future directions in Section~\ref{conc}.  Several complicated explicit results are given in an attached {\tt Mathematica} notebook.

\section{$\mathcal{N}=4$ stress-tensor four-point function}
\label{4point}

The main object of study in this work is the stress tensor multiplet four-point function. We begin by discussing general constraints on these correlators coming from invariance under the $\mathcal{N}=4$ superconformal algebra. We then discuss the large $N$ strong coupling expansion in Mellin space for the ${\cal N} = 4$ SYM theory.  Finally, we discuss how to constrain the terms in this expansion from the known Type IIB S-matrix in the flat space limit, as well as using the ${\cal F}_4(\tau, \bar \tau)$ and ${\cal F}_2(\tau, \bar \tau)$ introduced in Eqs.~\eqref{calF1Def} and \eqref{calF2Def} in the Introduction.

\subsection{Setup}
\label{setup}

As mentioned in the Introduction, we denote the bottom component of the stress tensor multiplet by $S$.  This operator is a dimension 2 scalar in the ${\bf 20}'$ of the $SU(4)_R \cong SO(6)_R$, and can thus be represented as a rank-two traceless symmetric tensor $S_{IJ}(\vec{x})$, with indices $I, J = 1, \ldots, 6$.  However, in order to avoid a proliferation of indices, it is customary to contract them with null polarization vectors $Y^I$, with $Y \cdot Y = 0$.  Superconformal symmetry \cite{Dolan:2001tt} implies that the four-point function of $S(\vec{x}, Y) \equiv S_{IJ}(\vec{x}) Y^I Y^J$ takes the form
 \es{FourPoint}{
  \langle S(\vec{x}_1, Y_1) \cdots S(\vec{x}_4, Y_4) \rangle 
   = \frac{1}{\vec{x}_{12}^4 \vec{x}_{34}^4}
     \vec{\cal S} \cdot \vec{\cB} \,, \qquad
     \vec{\cal S} \equiv  \vec{\cS}_\text{free} + \vec{{\bf S}} {\cal T}  \,,
 }
where $\vec{x}_{ij}  \equiv \vec{x}_i - \vec{x}_j$, and where
 \es{SThetaB}{
  \vec{\cS}_\text{free} &\equiv \begin{pmatrix}
   1 & U^2 & \frac{U^2}{V^2} & \frac{1}{c} \frac{U^2}V & \frac 1c \frac UV  & \frac 1c U
  \end{pmatrix} \,, \\
   \vec{{\bf S}} &\equiv \begin{pmatrix}
    V & UV & U & U(U- V - 1) & 1 - U - V & V (V - U - 1) 
   \end{pmatrix} \,, \\
   {\cal B} &= \begin{pmatrix}
    Y_{12}^2 Y_{34}^2 & Y_{13}^2 Y_{24}^2 & Y_{14}^2 Y_{23}^2
     & Y_{13} Y_{14} Y_{23} Y_{24} & Y_{12} Y_{14} Y_{23} Y_{34}
      & Y_{12} Y_{13} Y_{24} Y_{34}
   \end{pmatrix} \,.
 }
Here, as before, $c$ is the conformal anomaly coefficient, which for an $SU(N)$ gauge group equals $c = (N^2 - 1)/4$;  the quantities $U \equiv \frac{ \vec{x}_{12}^2 \vec{x}_{34}^2}{ \vec{x}_{13}^2 \vec{x}_{24}^2}$ and $V\equiv \frac{ \vec{x}_{14}^2 \vec{x}_{23}^2}{ \vec{x}_{13}^2 \vec{x}_{24}^2}$ are the usual conformal invariant cross-ratios;  and $Y_{ij} \equiv Y_i \cdot Y_j$ are $SO(6)_R$ invariants.  Importantly, the only non-trivial information in the correlator \eqref{FourPoint} is encoded in the single function $\cT(U, V)$.

\subsection{Strong coupling expansion}
\label{strong0}

We now restrict our discussion to the case of the $SU(N)$ ${\cal N} = 4$ SYM theory, and discuss the strong coupling 't Hooft limit, where we take $N\to\infty$ (or $c\to\infty$) with $\lambda\equiv g_\text{YM}^2N$ fixed.   If we further take $\lambda\to\infty$, the holographic correlator can be computed from Witten diagrams in an expansion around $AdS_5\times S^5$ supergravity.  In the strong coupling limit, it is convenient to work with the Mellin transform $\cM$ of $\cT$ via\footnote{The Mellin transform can also be defined away from the strong coupling limit.  For recent work on this topic, see \cite{Penedones:2019tng}.}
 \es{MellinDef}{
  \cT(U, V)
   = \int_{-i \infty}^{i \infty} \frac{ds\, dt}{(4 \pi i)^2} U^{\frac s2} V^{\frac t2 - 2}
    \Gamma \left[2 - \frac s2 \right]^2 \Gamma \left[2 - \frac t2 \right]^2 \Gamma \left[2 - \frac u2 \right]^2
    \cM(s, t) \,,
 } 
where $u \equiv 4 - s - t$.

Crossing symmetry $\cM(s, t) = \cM(t, s) = \cM(s, u)$ and the analytic properties of the Mellin amplitude (for a detailed description, see \cite{Chester:2019pvm}) then restrict $\cM(s, t)$ to have a $1/c$ and $1/\lambda$ expansion of the form
 \es{MIntro}{
   \cM =&\frac1 c\left[8\cM^\text{SG}+\lambda^{-\frac32}B_0^0 \cM^{0}  +\lambda^{-\frac52}\left[B_{2}^2 \cM^{2}+B_0^2\cM^{0}\right] +\lambda^{-3}\left[B_{3}^3 \cM^{3}+B_2^3\cM^{2}+B_0^3\cM^{0}\right]+O(\lambda^{-\frac{7}{2}})\right]\\
   &+\frac{1}{c^2}
   \left[\lambda^{\frac12} \overline{B_0^0} \cM^{0}+\left[\cM^{\text{SG}|\text{SG}}+\overline{B}^{\text{SG}|\text{SG}}_0\cM^0\right]+O(\lambda^{-1})\right]\\
   &+\frac{1}{c^3}
   \left[\lambda^{\frac32} \overline{\overline{B_2^2}} \cM^{2} +O(\lambda^{1})\right]+\frac{1}{c^4}
   \left[\lambda^3\left[\overline{\overline{\overline{B_3^3}}} \cM^{3}+\overline{\overline{\overline{B_2^3}}}\cM^{2}+\overline{\overline{\overline{B_0^3}}}\cM^{0}\right]+O(\lambda^{\frac52})\right]+O(c^{-5})\,,
}
which can be transformed to position space using \eqref{MellinDef} to get
 \es{TIntro}{
   \cT =&\frac1 c\left[8\cT^\text{SG}+\lambda^{-\frac32}B_0^0 \cT^{0}  +\lambda^{-\frac52}\left[B_{2}^2 \cT^{2}+B_0^2\cT^{0}\right] +\lambda^{-3}\left[B_{3}^3 \cT^{3}+B_2^3\cT^{2}+B_0^3\cT^{0}\right]+O(\lambda^{-\frac{7}{2}})\right]\\
   &+\frac{1}{c^2}
   \left[\lambda^{\frac12} \overline{B_0^0} \cT^{0}+\left[\cT^{\text{SG}|\text{SG}}+\overline{B}^{\text{SG}|\text{SG}}_0\cT^0\right]+O(\lambda^{-1})\right]\\
   &+\frac{1}{c^3}
   \left[\lambda^{\frac32} \overline{\overline{B_2^2}} \cT^{2} +O(\lambda^{1})\right]+\frac{1}{c^4}
   \left[\lambda^3\left[\overline{\overline{\overline{B_3^3}}} \cT^{3}+\overline{\overline{\overline{B_2^3}}}\cT^{2}+\overline{\overline{\overline{B_0^3}}}\cT^{0}\right]+O(\lambda^{\frac52})\right]+O(c^{-5})\,.
}
Here, the $B$'s are numerical coefficients that cannot be fixed from symmetry alone. As mentioned in the Introduction, terms at order $1/c^{g+1}$ correspond in the flat space limit to genus-$g$ corrections to the Type IIB S-matrix.  On $AdS_5\times S^5$, these terms receive contributions from $l$-loop Witten diagrams with $l\leq g$. The leading order term is tree-level supergravity, whose expression in Mellin and position space is \cite{Goncalves:2014ffa,Arutyunov:2000py}
\es{SintM}{
\cM^\text{SG}=&\frac{1}{(s-2)(t-2)(u-2)}\,,\qquad \mathcal{T}^\text{SG}= -\frac18U^2\bar D_{2,4,2,2}(U,V)\,,
}
where the position space expression is written in terms of the functions $\bar D_{r_1,r_2,r_3,r_3}(U,V)$ defined in \cite{Eden:2000bk}. The coefficient of $\cM^\text{SG}$ is fixed by requiring that the unprotected R-symmetry singlet of dimension two that appears in the conformal block decomposition of the free part $\vec{\cS}_\text{free}$ is not present in the full correlator \cite{Rastelli:2017udc}.  In our conventions \cite{Binder:2019jwn}, this amounts to setting the coefficient of $\mathcal{M}^\text{SG}$ to $8/c$. To the order considered in \eqref{TIntro}, the only loop term is $\mathcal{T}^{\text{SG}|\text{SG}}$, which arises from a loop Witten diagram with two supergravity vertices and so scales like $1/c^2$. This term was determined in \cite{Aprile:2017bgs,Alday:2017vkk,Alday:2018kkw} using unitarity methods up to a contact term ambiguity, which was further fixed in \cite{Chester:2019pvm}. Our convention for $\mathcal{M}^{\text{SG}|\text{SG}}$ follows \cite{Chester:2019pvm}, which is the Mellin transform of  $\mathcal{T}^{\text{SG}|\text{SG}}$ in \cite{Aprile:2017bgs}, although we will not make use of the explicit forms of these quantities.  

The remaining terms in \eqref{TIntro} arise from contact Witten diagrams whose vertices are higher derivative corrections to tree-level supergravity.  In particular, the functions $\cM^n$ and $\cT^n$ correspond to vertices of the form $D^{2n}R^4$, and their expressions in Mellin and position space are \cite{Alday:2014tsa}
 \es{treePoly}{
\cM^0 &= 1\,,\qquad\qquad\qquad\;\;  \mathcal{T}^0=U^2\bar D_{4,4,4,4}\,,\\
 \cM^{2} &= s^2+t^2+u^2 \,,\qquad  \mathcal{T}^2= 4U^2\left((1+U+V)\bar D_{5,5,5,5}-4\bar D_{4,4,4,4}\right)\,,\\
  \cM^{3} &= stu \,,\qquad\qquad \quad\;\;\, \mathcal{T}^3=- 8U^2\left(\bar D_{5,5,5,7}+(1+U+V)\bar D_{5,5,5,5}-8\bar D_{4,4,4,4}\right)\,.
}
We will now fix the various $B$'s in \eqref{TIntro} using type IIB string theory and/or the localization constraints. 

\subsection{Constraints from flat space type IIB string theory}
\label{flatCon}

Following the general approach of \cite{Penedones:2010ue}, one can relate the holographic correlator $\langle SSSS\rangle$ on $AdS_5\times S^5$ as written in terms of the Mellin amplitude $\mathcal{M}(s,t)$ in \eqref{MellinDef} to the type IIB S-matrix.  The scattering amplitude of four gravitons (or superpartners) in type IIB string theory takes the form
 \es{AGen}{
  \cA = \cA_\text{SG} f(s, t) \,,
 }
where $\mathcal{A}_\text{SG}$ is the tree-level supergravity amplitude, and $s,t,u=-s-t$ are the Mandelstam invariants.  The full amplitude as well as the tree-level supergravity amplitude in \eqref{AGen} depend on the momenta and polarizations of the scattered particles, which is information that we suppress in writing down \eqref{AGen}.  The function $f(s, t)$ has been computed in a small $g^2_s$ expansion to genus-two for finite $\ell_s$ \cite{Gomez:2010ad,DHoker:2005jhf}, and to genus-three \cite{Gomez:2013sla} to the lowest few orders in $\ell_s$. We will consider the following terms in the small $g_s$ and $\ell_s$ expansion:
 \es{A}{
   f(s, t) =& \left[\left(1+\ell_s^6f^0_{R^4}(s,t)+\ell_s^{10}f^0_{D^4R^4}(s,t)+\ell_s^{12}f^2_{D^6R^4}+O(\ell_s^{14}) \right)\right.\\
&\left.+g_s^2\left( \ell_s^6f^1_{R^4}(s,t)+\ell_s^8f^1_{\text{SG}|\text{SG}}(s,t)+\ell_s^{10}f^1_{D^4R^4}(s,t)+\ell_s^{12}f^2_{D^6R^4}+O(\ell_s^{14}) \right)\right.\\
&\left.+g_s^4\left(\ell_s^6f^2_{R^4}(s,t)+\ell_s^{10}f^2_{D^4R^4}+\ell_s^{12}f^2_{D^6R^4}+O(\ell_s^{14}) \right)\right.\\
&\left.+g_s^6\left(\ell_s^6f^3_{R^4}(s,t)+\ell_s^{10}f^3_{D^4R^4}+\ell_s^{12}f^3_{D^6R^4}+O(\ell_s^{14}) \right)+O(g_s^8)\right]\,.
}
Higher orders in $\ell_s$ can come from contact terms of higher derivative correction to supergravity, which are analytic in $s,t,u$ and have an expansion in $g_s$, as well as loops, which are non-analytic in $s,t,u$. The first few higher derivative terms are $R^4$, $D^4R^4$, and $D^6R^4$. These are the only protected terms.  They receive corrections at genus-zero for $R^4$, genus two for $D^4R^4$, and up to genus three for $D^6R^4$. These take the form
\es{fs}{
&f_{R^4}^0=\frac{\zeta(3)}{32}stu\,,\qquad f^0_{D^4R^4}=\frac{\zeta(5)}{2^{10}}stu(s^2+t^2+u^2)\,,\qquad f^0_{D^6R^4}=\frac{\zeta(3)^2}{2^{11}}(stu)^2\,,\\
& f^1_{R^4}=\frac{\pi^2}{96}stu\,,\qquad\;\; \;f^1_{D^4R^4}=0\,,\qquad\qquad\qquad\qquad\qquad\;\;\, f^1_{D^6R^4}=\frac{\pi^2\zeta(3)}{3\cdot2^{11}}(stu)^2\,,\\
& f^2_{R^4}=0\,,\qquad\qquad\quad f^2_{D^4R^4}=\frac{\pi^4}{2^{9}\cdot135}stu(s^2+t^2+u^2)\,,\;\, \,f^2_{D^6R^4}=\frac{\pi^4}{15\cdot2^{11}}(stu)^2\,,\\
& f^3_{R^4}=0\,,\qquad\qquad\quad f^3_{D^4R^4}=0\,,\qquad\qquad\qquad\qquad\qquad\;\;\, f^3_{D^6R^4}=\frac{\pi^6}{8505\cdot2^{10}}(stu)^2\,.\\
}
The only loop term shown in \eqref{A} is the one-loop term with two supergravity vertices, which can be computed from the genus-zero supergravity term using unitarity cuts \cite{Green:2008uj}.

The Mellin amplitude $\cM(s,t)$ is then related to the function $f(s, t)$ according to the flat space limit formula \cite{Penedones:2010ue,Fitzpatrick:2011hu,Chester:2018aca,Binder:2018yvd,Binder:2019jwn}:
 \es{Gotfst}{
   f(s, t) = \frac{stu}{2048\pi^2g_s^2\ell_s^8 }\lim_{L/\ell_s \to \infty} L^{14} \int_{\kappa-i\infty}^{\kappa+ i \infty} \frac{d\alpha}{2 \pi i} \, e^\alpha \alpha^{-6} {\cal M} \left( \frac{L^2}{2 \alpha} s, \frac{L^2}{2 \alpha} t \right) \,,
 }
where the momenta of the flat space S-matrix are restricted to lie within five of the ten dimensions. (When taking this limit, one uses the AdS/CFT dictionary\footnote{In the strong coupling limit we consider in this paper, the $\theta$ angle does not appear.}
 \es{dict}{
 \frac{L^4}{\ell^4_{s}}= \lambda = {g_\text{YM}^2 N}\,, \qquad
  g_s = \frac{g_\text{YM}^2}{4\pi} 
 }
to first write the correlation function in terms of $g_s$, $\ell_s$, and $L$, and then one takes $L / \ell_s$ to infinity as in \eqref{Gotfst}.)  We can then use the known terms in \eqref{fs} for the type IIB S-matrix to fix the leading $s,t$ terms $B_m^m$ in the $AdS_5\times S^5$ correlator:
 \es{fixFlat}{
  &\text{\underline{Constraints from flat space limit:}} \\
R^4:&\qquad B^0_0=120\zeta(3)\,,\qquad \overline{B_0^0}=\frac58\,,\\
D^4R^4:&\qquad B_2^2={630\zeta(5)}\,,\qquad \overline{\overline{B_2^2}}=\frac{7}{3072}\,,\\
D^6R^4:&\qquad B_3^3={5040\zeta(3)^2}\,,\quad\, \overline{\overline{\overline{B_3^3}}}=\frac{1}{221184}\,,\\
}
where the constraints on the $R^4$ and $D^4R^4$ coefficients were already derived in this way in \cite{Goncalves:2014ffa}. Note that the $R^4$ term is thus entirely fixed from the flat space limit alone. 

\subsection{Constraints from supersymmetric localization}
\label{SUSYCon}

As mentioned in the Introduction, we can also constrain $\langle SSSS\rangle$ just from the mass-deformed sphere free energy $F(m,\tau, \bar \tau)$, which Ref.~\cite{Pestun:2007rz} expressed as an $N$-dimensional matrix model integral using supersymmetric localization.  We have two such constraints, one coming from $\cF_2(\tau, \bar \tau) \equiv \partial_\tau\partial_{\bar\tau}\partial_m^2F\vert_{m=0}$ and one from $\cF_4(\tau, \bar \tau) \equiv \partial_m^4F(m,\tau, \bar \tau)\big\vert_{m=0}$.  The first one was shown in \cite{Binder:2019jwn} to take the form\footnote{Note that for the $SU(N)$ $\cN=4$ SYM theory, we have $ \partial_\tau \partial_{\bar \tau} F = - \frac{c \lambda^2}{32 \pi^2 N^2} =- \frac{c \lambda^2}{32 \pi^2 (4c + 1)} $.}
 \es{ConstraintOld}{
   \frac{ \cF_2(\tau, \bar \tau)}{ \partial_\tau \partial_{\bar \tau} F}= \frac{32 c}{\pi} \int dr \, d\theta\,  r^3 \sin^2 \theta
    \frac{r^2 - 1- 2 r^2 \log r}{(r^2 - 1)^2} \frac{\cT(1 + r^2 - 2r \cos \theta, r^2)}{(1 + r^2 - 2r \cos \theta)^2}  \,.
 }
As we will show in the next section, we can simplify this expression using crossing symmetry, obtaining 
  \es{constraint1}{
\frac{\cF_2(\tau, \bar \tau)}{\partial_\tau\partial_{\bar\tau}F}&=  8 c I_2[\cT]  \,,\qquad
I_2[\cT]\equiv -\frac{2}{\pi}\int drd\theta\frac{r^3\sin^2\theta}{U^2}\cT(U, V) \bigg|_{\substack{U = 1 + r^2 - 2 r \cos \theta \\
    V = r^2}}\,.
 }
The second constraint, whose detailed derivation we postpone until Section~\ref{LocCon}, takes the form
     \es{d4FAgainSec2}{
   \cF_4 &= -48 \zeta(3) c
   -c^2 I_4[\cT]\,,\\
     I_4[\cT]&\equiv\frac{32}{\pi}  \int dr\, d\theta\, r^3 \sin^2 \theta \, \frac{1 + U + V}{U^2} \bar{D}_{1,1,1,1}(U,V) \cT(U, V) \bigg|_{\substack{U = 1 + r^2 - 2 r \cos \theta \\
    V = r^2}}\,.
 }
 
The right-hand sides of Eqs.~\eqref{constraint1}--\eqref{d4FAgainSec2} involve the integrals $I_2[\mathcal{G}(U,V)]$ and $I_4[\mathcal{G}(U,V)]$, respectively, which, when evaluated on the functions of position defined in \eqref{SintM}--\eqref{treePoly} are
\es{integrals}{
I_2[\cT^\text{SG}]&=\frac{1}{32}\,,\quad\qquad I_2[\cT^0]=-\frac{1}{40}\,,\qquad I_2[\cT^2]=-\frac{2}{35}\,,\qquad I_2[\cT^3]=-\frac{4}{35}\,,\\
I_4[\cT^\text{SG}]&=3-6\zeta(3)\,,\quad I_4[\cT^0]=\frac{16}{5}\,,\qquad I_4[\cT^2]=\frac{272}{35}\,,\qquad I_4[\cT^3]=\frac{512}{35}\,.\\
}
The $I_2[\cT^\text{SG}]$, $I_2[\cT^0]$, and $I_2[\cT^2]$ integrals were first computed numerically in \cite{Binder:2019jwn} using the position space expressions \eqref{treePoly}. In Appendix~\ref{ImtauAn}, we confirm these results analytically using the simplified version \eqref{constraint1} of the constraint \eqref{ConstraintOld}, and we also compute analytically the new integral $I_2[\cT^3]$. The integrals for $I_4[\mathcal{G}]$ quoted in \eqref{integrals} were all computed numerically to high precision.  It would be interesting to develop an analytical method for computing them.

The LHS of \eqref{constraint1} was computed to leading order in the 't Hooft limit in~\cite{Russo:2013kea}, and used along with the flat space limit in~\cite{Binder:2019jwn} to fix the coefficients of both $R^4$ and $D^4R^4$ at genus zero.  In \cite{Chester:2019pvm}, this computation was extended to $O(N^{-6})$ and finite $\lambda$ using topological recursion.  As reviewed in Appendix~\ref{topApp}, the finite $\lambda$ result takes the form of a single Fourier integral, which can then be expanded analytically to any order in $1/\lambda$ following Appendix~D of~\cite{Binder:2019jwn}.  This can be used to fix the remaining non-zero genus terms in $R^4$ and $D^4R^4$, as well as the one-loop ambiguity term $\cM^0$, giving
\es{oldBs}{
 &\text{\underline{Constraints from $\cF_2$:}} \\
R^4:&\qquad B^0_0=120\zeta(3)\,,\qquad \overline{B_0^0}=\frac58\,,\\
\text{SG}|\text{SG}:&\qquad \overline{B}^{\text{SG}|\text{SG}}_0=\frac{15}{4}\,,\\
D^4R^4:&\qquad 16 B_2^2 + 7 B_0^2 =-3150\zeta(5)\,,\qquad   16 \overline{\overline{B_2^2}} + 7\overline{\overline{B_0^2}} =-\frac{35}{3072} \,. 
}

In Appendix~\ref{topApp}, we similarly use topological recursion to compute the LHS of \eqref{d4FAgainSec2} to $O(N^{-6})$ for finite $\lambda$.\footnote{We also use the method of orthogonal polynomials in Appendix \ref{topApp} to compute the non-instanton terms in $\cF_4 (\tau, \bar \tau)$ for finite $N$ and $\lambda$, following a similar calculation in \cite{Chester:2019pvm} for $\cF_2 (\tau, \bar \tau)$. While this result is not directly applicable to the strong coupling expansion considered here, it can be used to check the topological recursion expression for large $N$.} The result now involves two Fourier integrals, which cannot be analytically expanded in $1/\lambda$ as in Appendix~D of~\cite{Binder:2019jwn} unless the Fourier integrals factorize. Instead, we had to resort to a numerical large $\lambda$ expansion, which we show in \eqref{Ffinal}. Without using the flat space limit, the two integrated constraints can then be used to fix more coefficients beyond those in \eqref{oldBs}, namely we completely determine the $D^4 R^4$ coefficients and determine relations between the $D^6 R^4$ ones:   
\es{fixLoc}{
&\text{\underline{Constraints from $\cF_4$ and $\cF_2$:}} \\
D^4R^4:&\qquad B_0^2=-3B_2^2=-1890\zeta(5)\,,\qquad  \overline{\overline{B_0^2}}=-3  \overline{\overline{B_2^2}}=-\frac{7}{1024} \,, \\
D^6R^4:&\qquad 630\zeta(3)^2=\frac{7B_0^3}{32}+{B_3^3}\,,\qquad {{B_2^3}}=-1260\zeta(3)^2\,,\\
&\qquad\frac{1}{1769472}=\frac{7\overline{\overline{\overline{B_0^3}}}}{32}+{\overline{\overline{\overline{B_3^3}}}}\,,\qquad \overline{\overline{\overline{B_2^3}}}=-\frac{1}{884736}\,.
}
In addition, we can check that the constraint coming from $\cF_4$ by itself is sufficient to determine the $R^4$ coefficients constrained using $\cF_2$ in \eqref{oldBs}.  We cannot perform a similar check for $\overline{B}^{\text{SG}|\text{SG}}_0$ because we have not expanded $\cF_4$ to this order. Note that the localization constraints completely fix the $D^4R^4$ terms, which matches what we found from type IIB string theory in \eqref{fixFlat}. This is a nontrivial check of AdS/CFT at this order.

\subsection{$\langle SSSS\rangle$ to order $1/N^8$}

By combining the string theory and localization constraints we can fix all the terms in \eqref{MIntro} to get
\es{MellinFinal}{
   \cM =&\frac1 c\left[\frac{8}{(s-2)(t-2)(u-2)}+\frac{120\zeta(3)}{\lambda^{\frac32}}   +\frac{630\zeta(5)}{\lambda^{\frac52}}\left[s^2+t^2+u^2-3\right]\right.\\
   &\left.\qquad+\frac{5040\zeta(3)^2}{\lambda^{3}}\left[stu-\frac14(s^2+t^2+u^2)-4\right]+O(\lambda^{-3})\right]\\
   &+\frac{1}{c^2}
   \left[\frac{5\sqrt{\lambda}}{8}+\mathcal{M}^{\text{SG}|\text{SG}}+\frac{15}{4} +O(\lambda^{-\frac32})\right]\\
   &+\frac{1}{c^3}
   \left[\frac{7\lambda^{\frac32}}{3072}\left[s^2+t^2+u^2 - 3  \right]+O(\lambda)\right]\\
    &+\frac{1}{c^4}
   \left[\frac{\lambda^{3}}{221184} \left[stu-\frac14(s^2+t^2+u^2)-4\right]+O(\lambda^{\frac52})\right]+O(c^{-5})\,,
}
which is one of our main results, and which we also quoted in the Introduction.  In particular, we have fixed the genus-zero and genus-three terms in $D^6R^4$, where the latter scales as $\lambda^3/N^8$ and is the first result in $\mathcal{N}=4$ SYM at three orders beyond the planar limit! Note that we cannot yet fix the genus-one and genus-two terms in $D^6R^4$, since we have been unable to accurately expand the localization constraint at large $\lambda$ to the required orders yet.

Now that $\langle SSSS\rangle$ has been fixed to the order shown in \eqref{MellinFinal}, we can use it to extract any CFT data to this order that we like. For instance, we find the anomalous dimensions $\gamma_j$ of the unique lowest twist even spin $j$ double trace operators $[S\partial_{\mu_1}\dots\partial_{\mu_j}S]$ to be
\es{anom}{
\gamma_j=&\frac{1}{c}\left[-\frac{24}{(j+1)(j+6)}-\frac{4320\zeta(3)}{7\lambda^{\frac32}}\delta_{j,0}-\frac{\zeta(5)}{\lambda^{\frac52}}\left[{30600}\delta_{j,0}+\frac{201600}{11}\delta_{j,2}\right]\right.\\
&\left.\qquad+\frac{\zeta(3)^2}{\lambda^{3}}\frac{3628800}{11}\delta_{j,2}+O(\lambda^{-\frac72})\right]\\
&+\frac{1}{c^2}\left[-\frac{45\sqrt{\lambda}}{14}\delta_{j,0}+\frac{24 \left(7 j^4+74 j^3-553 j^2-4904 j-3444\right)}{(j-1) (j+1)^3 (j+6)^3
   (j+8)}-\frac{135}{7}\delta_{j,0}\right]\\
   &+\frac{1}{c^3}\left[-{\lambda^{\frac32}}\left[\frac{85}{768}\delta_{j,0}+\frac{35}{528}\delta_{j,2}\right]+O(\lambda)\right]+\frac{1}{c^4}\left[\frac{5{\lambda^{3}}}{16896}\delta_{j,2}+O(\lambda)\right]+O(c^{-5})\,,
}
where the three $O(c^{-1})$ terms were computed in \cite{Arutyunov:2000ku,DHoker:1999mic}, \cite{Goncalves:2014ffa}, and \cite{Binder:2019jwn}, respectively, while the one-loop supergravity term was computed in \cite{Aprile:2017bgs,Chester:2019pvm}. Contact terms with $n$-derivatives only contribute to operators up to spin $n/2-4$, as explained in \cite{Heemskerk:2009pn}. For higher twist there are many degenerate double trace operators, so one would need to compute many different half-BPS correlators to determine their anomalous dimensions \cite{Alday:2017xua,Aprile:2017bgs}.

\section{Constraints from the sphere partition function}
\label{LocCon}

In this section, let us complete our discussion by relating $\cT(U, V)$ defined in \eqref{FourPoint} to the derivatives of the $S^4$ partition function of the mass-deformed ${\cal N} = 4$ SYM theory that were introduced in Eqs~\eqref{calF1Def}--\eqref{calF2Def}. Both quantities involve special types of supersymmetric operators on $S^4$ that were considered in \cite{Binder:2019jwn}.  The first type are specific components of $S_{IJ}$ that are Coulomb branch operators from an ${\cal N} = 2$ point of view, and that are placed at the poles of $S^4$.  Adding them to the action is equivalent to changing the gauge coupling \cite{Gerchkovitz:2014gta,Gomis:2014woa,Gerchkovitz:2016gxx}.  The second type of supersymmetric operators considered in \cite{Binder:2019jwn} are integrated operators that couple to an ${\cal N} = 2$-preserving real-mass deformation.  As mentioned in the Introduction, Ref.~\cite{Binder:2019jwn} considered only the quantity $\cF_2(\tau, \bar \tau) = \partial_\tau\partial_{\bar\tau} \partial_m^2 F\big\vert_{m=0}$ (defined in Eq.~\eqref{calF2Def}) that equals the four-point function of two operators of the first type mentioned above and two operators of the second type.  This quantity gave the constraint on $\cT$ given in \eqref{ConstraintOld}.    In this section we consider the constraint coming from the quantity $\cF_4(\tau, \bar\tau) \equiv \partial_m^4 F\big\vert_{m=0}$ (defined in Eq.~\eqref{calF1Def}).  This quantity corresponds to four integrated insertions, and, for reasons that will become clear, this integrated correlator is more difficult to analyze than the correlator that led to \eqref{ConstraintOld}.

\subsection{Correlators in ${\cal N} = 4$ SYM}

Each integrated insertion is a linear combination of specific components of $S_{IJ}$ and specific components of the $\Delta = 3$ operators $P_{AB}$ and their conjugates $\bar P^{AB}$ transforming in the ${\bf 10}$ and $\overline{\bf 10}$ of $SU(4)_R$, respectively.  Here, the indices $A, B = 1, \ldots, 4$ are placed in a lower (upper) position when they correspond to the ${\bf 4}$ ($\overline{\bf 4}$) of $SU(4)_R$.  Since the ${\bf 10}$ and $\overline{\bf 10}$ are rank-two symmetric products of  ${\bf 4}$ and $\overline{\bf 4}$, respectively, the operators $P_{AB}$ and $\bar P^{AB}$ are symmetric tensors. For concreteness, we normalize $S$ and $P$ such that
 \es{NormSP}{
  \langle S(\vec{x}_1, Y_1) S(\vec{x}_2, Y_2) \rangle
   = \frac{(Y_1 \cdot Y_2)^2}{\abs{\vec{x}_{12}}^4} \,, \qquad 
     \langle P(\vec{x}_1, \bar X_1) \bar P (\vec{x}_2, X_2) \rangle
   = \frac{(\bar X_1 \cdot  X_2)^2}{\abs{\vec{x}_{12}}^6} \,,
 }
where we wrote $P(\vec{x}, X) = P_{AB}(\vec{x}) \bar X^A \bar X^B$ and $\bar P(\vec{x}, \bar X) = \bar P^{AB}(\vec{x}) X_A  X_B$ using polarization vectors $\bar X$ and $X$, respectively.  The dot product in \eqref{NormSP} and in subsequent equations stands for contraction using the Kronecker delta symbol. Before discussing in detail which components of $S$, $P$, and $\bar P$ participate in the mass deformation, let us point out that the four-point function $\langle P \bar P P\bar P \rangle$ is restricted by conformal symmetry and R-symmetry to take the form
 \es{PPPP}{
  \langle P(\vec{x}_1, \bar X_1) \bar P(\vec{x}_2,  X_2)  P(\vec{x}_3, \bar X_3) \bar P(\vec{x}_4, X_4) \rangle
   = \frac{1}{\abs{\vec{x}_{12}}^6  \abs{\vec{x}_{34}}^6  } \vec{\cal P}(U, V) \cdot \vec{\cal B}_P \,, 
 }
where a basis for the three distinct $SU(4)_R$ invariants can be taken to be
 \es{BPDef}{
   \vec{\cal B}_P \equiv \begin{pmatrix}
    (\bar X_1 \cdot X_2)^2 (\bar X_3 \cdot X_4)^2 &  (\bar X_1 \cdot X_4)^2 (\bar X_3 \cdot X_2)^2 & 
    (\bar X_1 \cdot X_2) (\bar X_3 \cdot X_4)  (\bar X_1 \cdot X_4) (\bar X_3 \cdot X_2)
   \end{pmatrix} \,.
 }
Thus, the $\langle P \bar P P \bar P \rangle$ correlator involves three functions ${\cal P}_i(U, V)$, with $i=1, 2, 3$.  We will also need the mixed correlator $\langle SS \bar P P \rangle$, which also involves three functions that we denote by ${\cal R}_i(U, V)$, with $i = 1, 2, 3$:
 \es{SSPP}{
  \langle S(\vec{x}_1, Y_1) S(\vec{x}_2, Y_2)  \bar P(\vec{x}_3,  X_3)  P(\vec{x}_4, \bar X_4) \rangle
   = \frac{1}{\abs{\vec{x}_{12}}^4  \abs{\vec{x}_{34}}^6  } \vec{\cal R}(U, V) \cdot \vec{\cal B}_{SP} \,,
 }
where, in this case, the basis of $SU(4)_R$ invariants can be taken to be 
 \es{BSPDef}{
  \vec{\cal B}_{SP} \equiv \begin{pmatrix}
   (Y_1 \cdot Y_2)^2 ( X_3 \cdot \bar X_4)^2 & 
    (Y_1 \cdot Y_2) \{\bar  X_4, X_3, Y_1, Y_2 \} & 
    \{ \bar X_4,  X_3, Y_1, Y_2 \}^2
  \end{pmatrix} \,.
 }
Here $\{ \bar X,  X, Y_1, Y_2 \}$ is an $SU(4)_R$ invariant that can be formed in the product ${\bf 10} \otimes \overline{\bf 10} \otimes {\bf 20}' \otimes {\bf 20}'$ defined in Eq.~(A.8) of \cite{Binder:2019jwn}.  

The coefficients ${\cal P}_i(U, V)$ and ${\cal R}_i(U, V)$ are related by the supersymmetric Ward identity to the function ${\cal T}(U, V)$ appearing in the $\langle SSSS \rangle$ correlator.  These relations are tedious to derive and quite complicated, so they are relegated to Appendix~\ref{WARDAPPENDIX}\@.  Separating out the contribution from the free theory as in Eq.~\eqref{FourPoint}, these relations take the form
 \es{PRSchem}{
  \vec {\cal P}(U, V) &= \vec {\cal P}_\text{free}(U, V) + \vec {\bf P}(U, V, \partial_U, \partial_V) {\cal T}(U, V) \,, \\
  \vec {\cal R}(U, V) &= \vec {\cal R}_\text{free}(U, V) + \vec {\bf R}(U, V, \partial_U, \partial_V) {\cal T}(U, V) \,,
 }
where the free theory contributions are
 \es{FreeContribs}{
   \vec {\cal P}_\text{free}(U, V) = \begin{pmatrix}
  1 & \frac{U^3}{V^3} & \frac{U(1 - U - V)}{4 c V^2}
  \end{pmatrix}  \,,
  \qquad
  \vec {\cal R}_\text{free}(U, V) = \begin{pmatrix}
   1 & 0 & 0 
  \end{pmatrix} \,,
 }
and the differential operators $\vec{\bf R}$ and $\vec{\bf P}$ are given in \eqref{SSPPward} and \eqref{ppppward1}--\eqref{ppppward3}, respectively.

\subsection{${\cal N} = 2$-preserving mass deformation}

Let us now discuss the mass deformation on $S^4$ in more detail.  It is customary to describe the ${\cal N} = 4$ SYM theory as an ${\cal N} = 1$ gauge theory with three adjoint chiral multiplets with scalar and fermionic components that we will denote by $Z_i$ and $\chi_i$, respectively, where $i = 1, 2, 3$, and with canonical kinetic terms.   In the decomposition of the ${\cal N} = 4$ vector multiplet into an ${\cal N} = 2$ vector multiplet and hypermultiplet, we can consider the $Z_i$ and $\chi_i$ with $i=1, 2$ as forming the hypermultiplet, and $Z_3$ and $\chi_3$ as part of the ${\cal N} = 2$ vector multiplet.    The mass deformation we consider corresponds to giving a mass to the hypermultiplet fields.  On $S^4$, this mass deformation takes the form
 \es{MassS4}{
  S_m = \int d^4 \vec{x}\, \sqrt{g} \left[  m \left( i J + K \right) + m^2 L \right] \,,
 }
where we assumed that the radius of $S^4$ is set to one, and where the operators $J$, $K$, and $L$ are given by
 \es{JKLDefs}{
  J &\equiv \frac 12 \sum_{i=1}^2 \tr \left( Z_i^2 + \bar Z_i^2 \right) \,, \qquad
   K \equiv - \frac 12 \sum_{i=1}^2 \tr (\chi_i \sigma_2 \chi_i + \tilde \chi_i \sigma_2 \tilde \chi_i) \,, \\
   L &\equiv   \tr \left[ \abs{Z_1}^2 + \abs{Z_2}^2  \right] \,.
 }
While the operators $K$ and $L$ multiplying $m$ and $m^2$, respectively, are familiar from a flat-space mass deformation, the operator $J$ (with its coefficient being inversely proportional to the radius of $S^4$) is present here in order to preserve ${\cal N} = 2$ supersymmetry on $S^4$.   

The operators appearing in \eqref{JKLDefs} can be written in terms of specific components of operators with well-defined transformation properties under the $SU(4)_R$ symmetry.  In particular, $J$ can be written in terms of $S_{IJ}$ and $K$ can be written in terms of $P_{AB}$ and $\bar P^{AB}$:
 \es{JCK}{
  J &= N_J \left[ S_{11} + S_{22} - S_{44}  - S_{55} \right] \,, \\
  K &= N_K \left[ P_{11} + P_{22} + \bar P^{11}  + \bar P^{22} \right] \,,
 }
up to some normalization constants that we denoted by $N_J$ and $N_K$.  (We will not need a similar expression for $L$ in what follows.)   One can determine the normalization constants in \eqref{JCK} by computing the two-point functions of $J$ and $K$ using the explicit description \eqref{JKLDefs} and the fact that supersymmetry implies that these two-point functions are protected.  One finds \cite{Binder:2019jwn}
 \es{NormRels}{
  N_K^2 = 8 N_J^2 = \frac{c}{\pi^4} \,,
 }
where, as before, the anomaly coefficient equals $c = (N^2 - 1)/4$.

\subsection{Four mass derivatives}

The four-point function contribution to the fourth mass derivative of the free energy is
 \es{FourthMass}{
  -\cF_4 = \left \langle  \left( \int d^4 \vec{x}\, \sqrt{g} \left(  i J + K  \right) \right)^4 \right\rangle
   + \text{(2- and 3-pt function contributions)} \,,
 }
where  the $2$- and $3$-point function contributions we did not write down explicitly involve the operator $L$.\footnote{They are $-12 \left\langle \left( \int d^4 \vec{x}\, \sqrt{g} \left(  i J + K  \right) \right)^2  \left( \int d^4 \vec{x}\, \sqrt{g} L \right) \right \rangle + 12 \left\langle   \left( \int d^4 \vec{x}\, \sqrt{g} L \right)^2 \right \rangle$.}   Using \eqref{JCK} as well as the general form of the four-point functions in \eqref{PPPP}, \eqref{SSPP}, and \eqref{FourPoint}, we find 
 \es{FourthMassAgain}{
  -\cF_4 &= N_J^4  I_{2, 2}^4 \left[ 16 \left( {\cal S}_1 + {\cal S}_2 + {\cal S}_3 \right) + 4 ({\cal S}_4 + {\cal S}_5 + {\cal S}_6 ) \right]
   - N_J^2  N_K^2 I_{2, 3}^4 \left[96 {\cal R}_1 + 288 {\cal R}_3 \right] \\
    &{}+ N_K^4 I_{3, 3}^4 \left[ 24 ({\cal P}_1 + {\cal P}_2 ) + 12 {\cal P}_3 \right]
     + \text{(2- and 3-pt function contributions)} \,,
 }
where $I_{\Delta_A, \Delta_B}^d[{\cal G}]$ denotes the integrated correlator on $S^d$ of four operators of dimensions $(\Delta_A, \Delta_A, \Delta_B, \Delta_B)$, which was studied in detail in \cite{Binder:2018yvd}.  To write it down explicitly, first note that on $\R^d$, such a correlation function takes the form ${\cal G}(U, V) / (\abs{\vec{x}_{12}}^{2 \Delta_A}\abs{\vec{x}_{34}}^{2 \Delta_B})$.   On a round $S^d$ of unit radius parameterized in stereographic coordinates such that the line element is $ds^2 = \Omega(\vec{x})^2 d\vec{x}^2$, the analogous correlator is 
 \es{CorrSd}{
   \frac{ \left[ \Omega(\vec{x}_1) \Omega(\vec{x}_2) \right]^{ - \Delta_A} 
      \left[ \Omega(\vec{x}_3) \Omega(\vec{x}_4) \right]^{ - \Delta_B}}{ \abs{\vec{x}_{12}}^{2 \Delta_A}\abs{\vec{x}_{34}}^{2 \Delta_B}} {\cal G}(U, V) \,, \qquad
       \Omega(\vec{x}) = \frac{1}{1 + \frac{\vec{x}^2}{4}} \,.
 }
The integrated correlator on $S^d$ is then 
 \es{IDef}{
  I_{\Delta_A, \Delta_B}^d[{\cal G}] 
   \equiv
    \int \left( \prod_{i=1}^4 d^d \vec{x}_i \right) 
     \frac{\left[ \Omega(\vec{x}_1) \Omega(\vec{x}_2) \right]^{d - \Delta_A} 
      \left[ \Omega(\vec{x}_3) \Omega(\vec{x}_4) \right]^{d - \Delta_B}}
      { \abs{\vec{x}_{12}}^{2 \Delta_A} \abs{\vec{x}_{34}}^{2 \Delta_B}   } 
       {\cal G}(U, V)\,.
 }
The quantity \eqref{IDef} was evaluated in \cite{Binder:2018yvd}, where, for $d=4$, it was found that
 \es{GotI4}{
  I_{\Delta_A, \Delta_B}^4[{\cal G}] 
   &= \frac{2^{17 - 2 \Delta_A - 2 \Delta_B} \pi^7 \Gamma(6 - \Delta_A - \Delta_B)}{3 \Gamma(4 - \Delta_A)^2 \Gamma(4 - \Delta_B)^2}  \\
   &\times
    \int dr \, d\theta\, r^3 \sin^2 \theta\, 
     \biggl[
      \bar D_{4 - \Delta_A, 4 - \Delta_A, 4 - \Delta_B, 4 - \Delta_B}(U, V) \frac{{\cal G}(U, V)}{U^{\Delta_A}}
      \biggr]_{\substack{U = 1 + r^2 - 2 r \cos \theta\\ V = r^2}} \,.
 }
Here, the $\bar D_{r_1, r_1, r_2, r_2}$ function is related to a contact Witten diagram in $AdS_4$ of four scalar fields dual to operators of dimensions $r_1$, $r_1$, $r_2$, $r_2$.  While one can write explicit position-space expressions for the $\bar D$ functions we need, for our purposes, however, the most useful definition the $\bar D$ function is through the Mellin transform:
 \es{barDDef}{
  \bar D_{r_1, r_1, r_2, r_2}(U, V) \equiv
   \int \frac{ds\, dt}{(4 \pi i)^2} U^{ \frac s2} V^{\frac t2} 
    \Gamma\left( - \frac{s + 2(r_1 - r_2)}{2}  \right) \Gamma\left( -\frac s2 \right) \Gamma\left( - \frac t2 \right)^2
     \Gamma\left( \frac{s + t + 2 r_1}{2} \right)  \,.
 }
Note that directly plugging  \eqref{GotI4} into \eqref{FourthMassAgain} is problematic, because $I_{\Delta_A, \Delta_B}^4[{\cal G}]$ contains a factor of $\Gamma(6 - \Delta_A - \Delta_B)$, which diverges for the $I_{3, 3}^4$ terms from the second line of \eqref{FourthMassAgain}.  We will thus have to find a way to regularize this divergence. 

The expression \eqref{FourthMassAgain} can be split into two parts:  one that is independent of ${\cal T}(U, V)$ corresponding to the free theory, and one that is linear in ${\cal T}(U, V)$ and its derivatives:
 \es{d4mFSplit}{
  -\cF_4 = -( \partial_m^4 F)_\text{free} \big|_{m=0} -( \partial_m^4 F)_{\cal T} \big|_{m=0} \,.
 } 
The first term, $ -( \partial_m^4 F)_\text{free} \big|_{m=0}$ can be calculated by plugging $\vec {\cal S} =\vec {\cal S}_\text{free}$, $\vec {\cal R} =\vec {\cal R}_\text{free}$, $\vec {\cal P} =\vec {\cal P}_\text{free}$ defined in \eqref{SThetaB} and \eqref{FreeContribs} into \eqref{FourthMassAgain}.  When performing this calculation, there are various divergences that arise and one has to be careful to regularize them properly.  We will not do that here, and instead calculate $ -( \partial_m^4 F)_\text{free} \big|_{m=0}$ using a different method that avoids some of these complications.   This alternative method relies on the observation that 
 \es{dm4Ffree}{
  -( \partial_m^4 F)_\text{free} \big|_{m=0} = - 4c \, \partial_m^4 F_H(m) \big|_{m=0} \,,
 }
where $F_H(m)$ is the $S^4$ free energy of a hypermultiplet of mass $m$.  The relation \eqref{dm4Ffree} holds true because in the zero coupling limit, the $SU(N)$ SYM theory of central charge $c = (N^2 - 1)/4$ has $N^2 - 1 = 4c$ such hypermultiplets of mass $m$.
 
The massive hypermultiplet free energy $F_H(m)$ can be determined as follows.  We start by writing down the theory of a single hypermultiplet with scalar and fermionic components $(Z_i, \chi_i)$, as in \eqref{MassS4}.  The action is
 \es{ActionFree}{
  S_\text{free}
   &= \sum_{i}  \int d^4x\, \sqrt{g} \biggl[
    \abs{\partial_\mu Z_i}^2
     + \frac{im}{2} \left[ (Z_i)^2 + (\bar Z_i)^2  \right] 
      + (2 + m^2) \abs{Z_i}^2 \\
    &{}-\tilde \chi_i^{T} \sigma_2 \bar \sigma^\mu D_\mu \chi_i
   - \frac m2 (\chi_i \sigma_2 \chi_i + \tilde \chi_i \sigma_2 \tilde \chi_i) \biggr] \,.
 }
The path integral $Z_H(m) = \int DX e^{-S_\text{free}[X]}$, where $X$ denotes collectively the hypermultiplet fields, is a Gaussian integral that can be evaluated as a ratio of a fermionic determinant to a bosonic one.  Up to an overall $m$-independent normalization, this ratio is\footnote{The eigenvalues of the bosonic operator are $(n+1 + i m)(n+2 - i m)$ and $(n + 1 - im ) (n + 2 + im )$, with $n= 0, 1, 2, \ldots$, each with degeneracy $D_n = \frac 16 (n+1)(n+2) (2n + 3)$.  The eigenvalues of the fermionic operator are $n + 2 + i m$ $n + 2 - im $, each with degeneracy $\tilde D_n = \frac13 (n+1)(n+2)(n+3)$.}
 \es{ZHFree}{
  Z_H(m) = \frac{\prod_{n=0}^\infty \left[ (n+2)^2 + m^2 \right]^{(n+1)(n+2)(n+3)/3} }
   {\prod_{n=0}^\infty  \left[\left( (n+1)^2 + m^2 \right) \left( (n+2)^2 + m^2 \right) \right]^{(n+1)(n+2)(2n+3)/12} } \,.
 } 
This expression can be simplified and then regularized:
 \es{ZHFreeAgain}{
  Z_H(m)   = \frac{1}{\prod_{n=1}^\infty  (n^2 + m^2)^{\frac n2}} = \frac{1}{H(m)} \,, \qquad
   H(m) \equiv e^{-(1 + \gamma) m^2} G(1 + im) G(1 - im) \,,
 }
where $G$ is the Barnes G-function and $\gamma$ is the Euler-Mascheroni constant.  The normalization of $Z_H(m)$ in \eqref{ZHFreeAgain}  was chosen such that $Z_H(0) = 1$.  The function $H(m)$ appeared in the supersymmetric localization computation of  \cite{Pestun:2007rz}, and indeed, the result \eqref{ZHFreeAgain} can be also read off from \cite{Pestun:2007rz}.  The equation \eqref{ZHFreeAgain} is imprecise, however, partly because the regularization of \eqref{ZHFreeAgain} possesses ambiguities, and partly because we dropped an unambiguous overall coefficient that depends on the radius of the sphere, as required by the conformal anomaly.  The ambiguity in the free energy $F_H = - \log Z_H$ consists of additive terms of the form $A + B m^2$ where both $A$ and $B$ are sums of holomorphic and anti-holomorphic functions of the complexified coupling $\tau$.\footnote{A sign that such an ambiguity is present is the appearance of the Euler-Mascheroni constant $\gamma$ in \eqref{ZHFreeAgain}, which suggests that this expression was derived in a particular regularization scheme.}  Such ambiguities, as well as the unambiguous overall coefficient, drop out from the fourth mass derivative of $F_H$ that we consider here.  Note that although the overall factor $e^{-(1 + \gamma) m^2}$ in \eqref{ZHFreeAgain} can be removed by a change of regularization scheme, we will nevertheless keep it for later convenience.  Using \eqref{dm4Ffree} and the expression for $Z_H(m) = e^{-F_H(m)}$ in \eqref{ZHFreeAgain}, we find
 \es{dFFree}{
  -(\partial_m^4 F)_\text{free} \big|_{m=0} = 48 c\,  \zeta(3) \,.
 }

What remains to be done is to evaluate the $\cT$-dependent contribution $-(\partial_m^4 F)_{\cal T}$ in \eqref{d4mFSplit}.  From \eqref{FourthMass}, it can be written as
 \es{d4mFT}{
  -(\partial_m^4 F)_{\cal T} \big|_{m=0}
   &=  \frac{c^2}{16 \pi^8}  I_{2, 2}^4 \left[ (1 + U + V)^2 {\cal T} \right]+  \frac{12 c^2}{ \pi^8} I_{3, 3}^4 \left[ (2{\bf P}_1 + 2{\bf P}_2  +  {\bf P}_3){\cal T} \right]  \\
   &{}-  \frac{12 c^2}{ \pi^8} I_{2, 3}^4 \left[ ({\bf R}_1 + 3 {\bf R}_3){\cal T} \right] 
    + \text{(2- and 3-pt function contributions)}  \,,
 }
where the $2$-point and $3$-point function contributions here are the subset of the ones from \eqref{FourthMass} that were not accounted for in $4c$ copies of the free theory.  We will not write them in detail because, as we will discuss, we believe that the boundary terms from the integration by parts we will be performing shortly precisely cancels them.  Such a phenomenon was observed also in \cite{Binder:2018yvd} in 3d.

Let us study the first three terms in \eqref{d4mFT} separately, and let us aim to write them in the ``canonical form''
 \es{CanForm}{
  \cF[Q] &= 
   \int dr\, d\theta\, r^3 \sin^2 \theta
     \frac{(1 + U + V) \cT(U, V)}{U^2} \\
    &\times 
    \int \frac{ds \, dt}{(4 \pi i)^2} U^{\frac s2} V^{\frac t2} 
      \Gamma^2 \left( - \frac{s}{2} \right) \Gamma^2 \left( - \frac{t}{2} \right)
       \Gamma^2 \left( - \frac{u}{2} \right)  
        Q(s, t, u) 
         \bigg|_{\substack{U = 1 + r^2 - 2 r \cos \theta \\V = r^2}} \,,
 }
where $u =  - 2 - s - t$, where each term will have a different function $Q(s, t, u)$.     The expression \eqref{CanForm} is designed such that when $Q  = 1$, the second line equals $\bar D_{1, 1, 1, 1}(U, V)$, as can be seen from \eqref{barDDef}.

For the first term in \eqref{d4mFT}, we combine \eqref{GotI4} with \eqref{barDDef} and shift the integration variables as appropriate to obtain
 \es{FristRewrite}{
  I_{2, 2}^4 \left[  (1 + U + V)^2 {\cal T} \right]
   = \cF [Q_{SSSS} ]   \,, \qquad
    Q_{SSSS} = \frac{128 \pi^7}{3} (s^2 + t^2 + u^2) \,.
 }
For the second term, we plug \eqref{RSimp} into \eqref{GotI4}, and then we integrate by parts to put all the derivatives on the $\bar D_{2, 2, 1, 1}$ function.  We obtain
 \es{SecondRewriteTmp}{
    &I_{2, 3}^4 \left[ ({\bf R}_1 + 3 {\bf R}_3){\cal T} \right]
     = \frac{32}{3} \pi^7 \int dr \, d \theta\, r^3 \sin^2 \theta\, 
      \frac{V(1 + U + V)}{U^2} 
      \biggl[
       U^2 \partial_U^2 + 4 U \partial_U  \\
       &{}+ U(U + V-1) \partial_U \partial_V
       +  U V \partial_V^2  + (3 U + V - 1) \partial_V + 2 
      \biggr]
       \bar D_{2, 2, 1, 1}(U, V) \bigg|_{\substack{U = 1 + r^2 - 2 r \cos \theta\\ V = r^2}}  \,.
 }
Then, using the expression \eqref{barDDef} for the $\bar D_{2, 2, 1, 1}$ function in Mellin space, we can write \eqref{SecondRewriteTmp} in the canonical form \eqref{CanForm}:
 \es{SecondRewrite}{
     I_{2, 3}^4 \left[ ({\bf R}_1 + 3 {\bf R}_3){\cal T} \right]
   = \cF[Q_{SSPP}] \,, \qquad
    Q_{SSPP} = -\frac{8 \pi^7}{3} tu \,.
 } 
For the third term in \eqref{d4mFT}, we should follow a similar procedure.  Since the prefactor in \eqref{GotI4} diverges when $\Delta_A = \Delta_B = 3$, we should evaluate this quantity in a way that avoids this divergence.  This can be done by first considering $\Delta_A=3$ and $\Delta_B = 3 -\epsilon$, both in the prefactor and in the expression for the $\bar D$ function in \eqref{GotI4}.\footnote{We are grateful to Thomas Dumitrescu for extensive discussions about this issue.}   We then use the Ward identity \eqref{PSimp} (which, as will be justified shortly, should also hold for non-zero $\epsilon$), integrate by parts, and then take $\epsilon \to 0$.  This procedure gives
 \es{ThirdRewrite}{
   I_{3, 3}^4 \left[ (2{\bf P}_1 + 2{\bf P}_2  +  {\bf P}_3){\cal T} \right]
    = \cF[Q_{PPPP}] \,, \qquad
     Q_{PPPP} = \frac{2 \pi^7}{3} u (u-2) \,.
 } 
The fact that we can use the same Ward identity \eqref{PSimp} when $\epsilon \neq 0$ can be justified as follows. Instead of considering the four-point functions of operators from the stress tensor multiplet, we can consider four-point functions where the first two operators are from the stress-tensor multiplet and the last two are from other half-BPS multiplets whose superconformal primaries $S_p$ transform in the $[0p0]$ irrep of $SU(4)_R$ and have dimension $\Delta_p = p$.  (The stress tensor multiplet corresponds to $p=2$.)  This multiplet contains generalizations $P_p$ and $\bar P_p$ of the $P$ and $\bar P$ operators, respectively, which transform in the $[2 (p-2) 0]$ and $[0 (p-2) 2]$ of $SU(4)_R$, respectively, and have scaling dimensions $\Delta_p + 1$.  The form of any of the mixed correlation functions with two operators from the $p=2$ multiplet and two operators from a $p \neq 2$ multiplet is precisely the same as when $p=2$, except that there is an additional factor of $(Y_3 \cdot Y_4)^{p-2}$.  Moreover, the  Ward identity relations relating $\langle P \bar P P_p \bar P_p\rangle$ to $\langle S S S_p S_p \rangle$ is independent of $p$.   Analytically continuing $\langle P \bar P P_p \bar P_p \rangle$ in $p$ to $p = 2- \epsilon$ leads to $\Delta_B = 3 - \epsilon$ as above.

Combining the above results, we have
 \es{d4FCombined}{
  -(\partial_m^4 F)_{\cal T} \big|_{m=0}
   &=  \frac{c^2}{16 \pi^8} 
     \cF \left[ Q_{SSSS} + 192 (Q_{PPPP} - Q_{SSPP} )   \right] \,.
 }

\subsection{Simplification using crossing symmetry and final formula}

One can simplify the formula \eqref{d4FCombined} using crossing symmetry.  Crossing symmetry relies on the observation that under a simultaneous relabeling of the pairs $(\vec{x}_i, Y_i)$ in \eqref{FourPoint}, the four-point function should remain unchanged.  There are $24$ orderings of these four pairs, but some of them leave \eqref{FourPoint} manifestly invariant.  There are six that do not, and the impose the following crossing constraints on the function $\cT$:
 \es{TCrossing}{
  \cT\left(\frac UV, \frac 1V\right) &= \cT\left(\frac VU, \frac 1U\right) = V^2 \cT(U, V) \,, \qquad
   \cT\left(\frac 1U, \frac VU\right) = \cT(U, V)  \,, \\
   \cT\left(\frac 1V, \frac UV\right) &= \cT(V, U)  = \frac{V^2}{U^2} \cT(U, V) \,.
 }
Thus, we write \eqref{CanForm} in five other equivalent ways by simply sending $(U, V)$ to either $\left(\frac UV, \frac 1V\right)$, $\left(\frac VU, \frac 1U\right)$, $\left(\frac 1U, \frac VU\right)$, $\left(\frac 1V, \frac UV\right)$, or $(V, U)$ (along with corresponding changes in $(r, \eta)$), and then using the relations in \eqref{TCrossing}.  Averaging over these six possibilities (the original expression \eqref{CanForm} as well as the five expressions obtained as above), one obtains a similar expression to \eqref{CanForm}, with the only difference being that the factor $U^{\frac s2} V^{\frac t2}$ is symmetrized in $s$, $t$, and $u$:
 \es{UReplacement}{
  U^{\frac s2} V^{\frac t2} \to 
   \frac{U^{\frac s2} V^{\frac t2} + 
    U^{\frac s2} V^{\frac u2} + U^{\frac u2} V^{\frac t2} + U^{\frac t2} V^{\frac s2}
     + U^{\frac t2} V^{\frac u2} + U^{\frac s2} V^{\frac u2}}{6}  \,,
 }
where $u = -2 - s- t$.  One can then rename $s$, $t$, and $u$ to rewrite \eqref{CanForm} such that $Q(s, t)$ is replaced by the symmetrized expression
 \es{QSym}{
  Q(s, t) \to \frac{Q(s, t) + Q(s, u) + Q(u, t) + Q(t, s) + Q(t, u) + Q(s, u)}{6} \,.
 }
After this symmetrization, Eqs.~\eqref{FristRewrite}, \eqref{SecondRewrite}, and \eqref{ThirdRewrite} imply that we can replace $Q_{SSSS}$, $Q_{SSPP}$, and $Q_{PPPP}$ by
 \es{QAfterSym}{
  Q_{SSSS} &= \frac{128 \pi^7}{3} (s^2 + t^2 + u^2) \,, \\
   Q_{SSPP} &= -\frac{8 \pi^7}{9} (st + su + tu)
    =  \frac{4 \pi^7}{9} ( s^2 + t^2 + u^2- 4)\,, \\
   Q_{PPPP} &= \frac{2 \pi^7}{9} (s^2 + t^2 + u^2 +4) \,.
 }
Plugging \eqref{QAfterSym} into \eqref{d4FCombined}, we see quite nicely that the dependence on $s$, $t$, and $u$ inside the argument of $\cF$ disappears, and we simply have
 \es{d4FCombinedAgain}{
  -(\partial_m^4 F)_{\cal T} \big|_{m=0}
   &=  \frac{32c^2}{ \pi} 
     \cF \left[ 1 \right] \,.
 }
Recalling that evaluating $\cF$ at $Q = 1$ means replacing the second line of \eqref{CanForm} with $\bar D_{1, 1,1 ,1}(U, V)$, we then conclude that
    \es{d4FAgain}{
  -(\partial_m^4 F)_{\cal T} \big|_{m=0} &= c^2 I_4[\cT]\,,
 }
with $I_4[\cT]$ defined in \eqref{d4FAgainSec2}.   Combining this expression with \eqref{dFFree}, we immediately obtain \eqref{d4FAgainSec2}.  This is our final formula for  $\cF_4(\tau, \bar \tau)$.

There are two loose ends to be tied up.  The first concerns the $2$- and $3$-point function contributions in \eqref{d4mFT}.  We note that the mass parameter $m$ has dimension $1$, so $m$ must couple to an operator of dimension precisely $3$ and $m^2$ must couple to an operator of dimension precisely $2$.  Away from zero coupling, the only such operators present are operators in the $SU(N)$ $\cN = 4$ stress tensor multiplet, whose canonically-normalized $2$- and $3$-point functions are proportional to $c$.  Thus, the entire $2$- and $3$-point function contribution to \eqref{FourthMass} must be proportional to $c$ as well, so any such contribution that we did not take into account would simply modify the first term in \eqref{d4FAgainSec2}.  However, we did check that with the formula \eqref{d4FAgainSec2} as written, there is agreement between the leading large $c$ supersymmetric localization result for $\cF_4$ and the explicit evaluation of the RHS of \eqref{d4FAgainSec2} using the known supergravity amplitude \eqref{SintM}.  This is a strong check that the coefficient of the term proportional to $c$ in  \eqref{d4FAgainSec2} is as written.

The second loose end concerns the simplified formula \eqref{constraint1} for the mixed derivative $\cF_2(\tau, \bar \tau)$.  As mentioned before, Ref.~\cite{Binder:2019jwn} derived the relation \eqref{ConstraintOld} between the mixed derivatives of the $\cN = 2^*$ partition function and an integral of $\cT(U, V)$.  One can simplify this formula using crossing symmetry, namely by taking the average between \eqref{ConstraintOld} and the expression obtained after replacing $(U, V)$ in \eqref{ConstraintOld} by $\left(\frac UV, \frac 1V\right)$, $\left(\frac VU, \frac 1U\right)$, $\left(\frac 1U, \frac VU\right)$, $\left(\frac 1V, \frac UV\right)$, or $(V, U)$, and after using \eqref{TCrossing} to write everything in terms of $\cT(U, V)$.  This procedure gives \eqref{constraint1}.

As a final comment, let us note that the integrated relations \eqref{constraint1} and \eqref{d4FAgainSec2} do not apply only to the $\cN = 4$ SYM theory with gauge group $SU(N)$.  These expressions apply equally well to $\cN = 4$ SYM with some other semi-simple gauge group $G$, for which $c = (\text{dim}\, G)/4$, where $\text{dim}\, G$ denotes the dimension of $G$.

\section{Discussion}
\label{conc}

The main result of this work was a new exact relation between four derivatives of the mass deformed sphere free energy, $\cF_4(\tau, \bar \tau) \equiv \partial_m^4F(m,\tau, \bar \tau)\big\vert_{m=0}$, and an integral of the four point function $\langle SSSS\rangle$ of the superconformal primary $S$ of the stress tensor multiplet in $\mathcal{N}=4$ SYM theory.   For gauge group $SU(N)$, we applied this constraint in the strong coupling 't Hooft limit at large $c\sim N^2$ and large $\lambda$, where the $\cN = 4$ SYM theory  is holographically dual to type IIB string theory on $AdS_5\times S^5$.  In combination with the constraint coming from $\cF_2(\tau, \bar \tau) =  \partial_\tau \partial_{\bar \tau} \partial_m^2F(m,\tau, \bar \tau)\big\vert_{m=0}$ derived in \cite{Binder:2019jwn}, the $\cF_4(\tau, \bar \tau) $ constraint allowed us to completely fix the $D^4R^4$ contributions to the $\langle SSSS\rangle$ correlation function.  (This contact interaction vertex contributes non-trivially at genus zero and genus two.)  In the flat space limit, we matched these contributions to the known $D^4R^4$ terms in the Type IIB S-matrix.  Using the constraint on $\langle SSSS \rangle$ from the known flat space S-matrix combined with the two constraints from $\cF_2$ and $\cF_2$, we were able to further fix the genus-zero and genus-three $D^6R^4$ term in $\langle SSSS\rangle$, where the latter scales as $\lambda^{3}c^{-4}$ and is the first known contribution to $\langle SSSS\rangle$ that has been computed at order $1/c^4$.

Looking ahead, it would be useful to develop an analytic large $\lambda$ expansion of $\cF_4(\tau, \bar \tau) $, as was achieved for $\cF_2(\tau, \bar \tau)$ in \cite{Chester:2019pvm}. In the latter case, the large $N$ and finite $\lambda$ expressions obtained by topological recursion were given in terms of a single Fourier integral, which could then be analytically expanded to any order in $\lambda$ using the method described in Appendix D of \cite{Binder:2019jwn}. For $\cF_4(\tau, \bar \tau) $, however, the large $N$ and finite $\lambda$ expressions that we derived in this work are given in terms of two Fourier integrals, which were not amenable to the method of Appendix D of \cite{Binder:2019jwn} unless the Fourier integrals factorized. Instead, we had to resort to a numerical large $\lambda$ expansion, which only gave precise results at low orders in $\lambda$. In particular, we were unable to compute the $N^0\lambda^0$, $N^0\lambda^{-\frac12}$, and $N^{-2}\lambda$ contributions to $\cF_4(\tau, \bar \tau) $. The first term could be used to confirm the derivation of the one-loop constant ambiguity $\overline{B}_0^{\text{SG}|\text{SG}}$ that was previously fixed in \cite{Chester:2019pvm} using $\cF_2(\tau, \bar \tau)$. The $N^0\lambda^{-\frac12}$ and $N^{-2}\lambda$ terms could be used to derive the genus one and two contributions, respectively, to $D^6R^4$ in $\langle SSSS\rangle$, which would complete the derivation of the $D^6R^4$ term begun in this work.

While in this work we considered the strong coupling 't Hooft limit, one could also consider the holographic limit where $N\to\infty$ and $\tau=\frac{\theta}{2\pi}+\frac{4\pi i}{g_\text{YM}^2}$ is finite.  In the flat-space limit, this strong coupling limit of the $\langle SSSS \rangle$ correlator should match the small $\ell_s$ and finite $\tau_s=\chi_s+ig_s^{-1}$ expansion of the type IIB S-matrix, where $\chi_s$ is the expectation value of the type IIB axion. The coefficients of the various powers of $1/c$ in $\langle SSSS \rangle$ for each expansion must be $SL(2,\Z)$ invariants of $\tau$ and $\bar \tau$.  In the flat space limit, the coefficients of $1/c^{7/4}$, $1/c^{9/4}$, and $1/c^{5/2}$ correspond to the protected $R^4$, $D^4R^4$, and $D^6R^4$ contact amplitudes that were derived in \cite{Green:1997as,Green:1998by,Green:1999pu,Green:2005ba,Green:2014yxa}. In \cite{Chester:2019jas}, the mixed mass derivative $\cF_2(\tau, \bar \tau)$ was studied in this limit, and combining the integrated constraints with the flat space limit, it was possible to completely determine the $\langle SSSS \rangle$ correlator at orders $1/c^{7/4}$ and $1/c^{9/4}$.  It would be very interesting, but much harder, to extend this analysis to $\cF_4(\tau, \bar \tau)$.  We leave this topic for future work \cite{CGPWW}.

In addition to the constraints on the $\langle SSSS \rangle$ correlator considered here, one could also consider new integrated constraints that come from derivatives in terms of the squashing parameter $b$ for the free energy $F(b,m,\tau, \bar \tau)$ on the squashed sphere, which was also computed in terms of a matrix model using localization in \cite{Hama:2012bg}. Of the three possible constraints:
\es{newb}{
\partial_b^4F(b,0,\tau)\big\vert_{b=1}\,,\qquad \partial_b^2\partial_m^2F(b,m,\tau)\big\vert_{b=1,m=0}\,,\qquad \partial_\tau\partial_{\bar\tau}\partial_b^2F(b,0,\tau)\big\vert_{b=1}\,,
}
we expect that only one of the first two constraints to be linearly independent from the two already considered, which is exactly enough constraints to fix the $D^6R^4$ term in $\langle SSSS\rangle$ purely from CFT\@. These three localization constraints could also be combined with the known type IIB S-matrix in the flat space limit to fix the four ambiguities in the one-loop term $\cM^{\text{SG}|R^4_\text{genus-0}}$ with one supergravity vertex and one genus-zero $R^4$ vertex \cite{Alday:2018pdi,Alday:2018kkw}, which scales like $\lambda^{-\frac32}N^{-4}$. One could similarly fix the $D^8R^4$ contact term to genus two.

Lastly, while the application of integrated constraints and localization to holographic correlators has been to the large $N$ expansion in this paper and previous work \cite{Binder:2019jwn,Chester:2019pvm,Chester:2019jas}, these relations are in fact non-perturbative, and so could be applied to the numerical bootstrap for $\mathcal{N}=4$ SYM \cite{Beem:2013qxa,Beem:2016wfs}. For this purpose, the finite $N$ formula for the perturbative part of the mass deformed free energy, as derived using orthogonal polynomials in Appendix~\ref{ortho2}, will be especially useful, especially if one could augment it with a similar formula for the contribution from the Nekrasov partition function. These constraints could allow one to impose the values of $\tau$ and $\bar\tau$ in the numerical bootstrap for finite $N$, just as $N$ was imposed in the original studies \cite{Beem:2013qxa,Beem:2016wfs} using the conformal anomaly $c$, and thereby solve $\mathcal{N}=4$ SYM numerically for all $\tau$, $\bar\tau$ and $N$.

\section*{Acknowledgments} 

We thank Thomas Dumitrescu and Ofer Aharony for useful discussions, as well as Damon Binder, Michael Green, Yifan Wang, and Congkao Wen for useful discussions and collaboration on related work.  SMC is supported by the Zuckerman STEM Leadership Fellowship. The work of SSP was supported in part by the US NSF under Grant No.~PHY-1820651 and by the Simons Foundation Grant No.~488653.  We thank the organizers of the workshops ``Bootstrap 2019'' and ``Scattering amplitudes and the conformal bootstrap''  which took place at the Perimeter Institute for Theoretical Physics and at the Aspen Center for Physics (ACP), respectively, for hospitality while this work was in progress. The ACP is supported by National Science Foundation Grant No.~PHY-1607611.

\appendix

\section{$\cF_4(\tau, \bar \tau)$ from supersymmetric localization}
\label{topApp}

In this appendix, we show how $\cF_4(\tau, \bar \tau)$ can be computed from the supersymmetric localization result of \cite{Pestun:2007rz}, following a similar calculation for $\cF_2(\tau, \bar \tau)$ in \cite{Chester:2019pvm}.  We will start by writing $\cF_4(\tau, \bar \tau)$ as an expectation value of an operator in a Gaussian matrix model. We then evaluate this expectation value to any order in $1/N^2$ using topological recursion \cite{Eynard:2004mh,Eynard:2008we}, or for finite $N$ and $\lambda = \frac{4\pi N}{\Im\tau}$ (ignoring instantons) using orthogonal polynomials \cite{mehta1981}.

\subsection{Matrix model expectation value}
\label{moreSetup}

As shown by Pestun \cite{Pestun:2007rz}, the $S^4$ partition function $Z=\exp(-F)$ of the $SU(N)$ ${\cal N} = 2^*$ is given by
\es{N2starMatrixModel}{
	Z(m, \tau, \bar \tau)
	= \int d^{N} a\,\delta\big(\sum_i a_i\big) e^{-\frac{8 \pi^2 N }{\lambda} \sum_i a_i^2} \abs{Z_\text{inst}(m, \tau)}^2 \frac{ \prod_{i < j}a_{ij}^2 H^2(a_{ij})}{ H(m)^{N-1} \prod_{i \neq j} H(a_{ij}+ m)} \,,
}
where we denoted $a_{ij}\equiv a_i-a_j$, and where the delta function enforces the $SU(N)$ constraint that the eigenvalues sum to zero.  The function $H(m)$ appearing in \eqref{N2starMatrixModel} was already defined in the main text in Eq.~\eqref{ZHFreeAgain}. The quantity $\abs{Z_\text{inst}(m, \tau)}^2$ represents the contribution to the localized partition function coming from instantons located at the North and South poles of $S^4$ \cite{Nekrasov:2002qd,Nekrasov:2003rj,Losev:1997tp,Moore:1997dj}, and can be ignored in the 't Hooft limit because it is non-perturbative when $g_\text{YM} \to 0$.

The LHS of the perturbative part of the integrated constraint \eqref{d4FAgain} is then
  \es{4m}{
- \partial_m^4F^\text{pert}\vert_{m=0}=&-12\zeta(3)+\sum_{i, j}\langle K'''(a_{ij})\rangle+3\sum_{i,j,k,l}\left[\langle K'(a_{ij})K'(a_{kl})\rangle-\langle K'(a_{ij})\rangle\langle K'(a_{kl})\rangle\right]\,,
 }
where $K(z)\equiv -\frac{H'(z)}{H(z)}$, and where the expectation values are taken in the Gaussian matrix model
\es{ZSUN}{
   Z= \int d^{N} a\,\delta\left(\sum_i a_i\right) e^{-\frac{8 \pi^2 N }{\lambda} \sum_i a_i^2}  \prod_{i < j}a_{ij}^2 \,.
 } 

The function $K'(z)$ can be simply expressed using its Fourier transform \cite{Russo:2013kea}
  \es{KFourier}{
 K'(z)=-\int_0^\infty d\omega\frac{2\omega[\cos(2\omega z)-1]}{\sinh^2\omega}\,.
 }
To calculate \eqref{4m}, we should then first compute the 2-body expectation value
\es{exp1}{
\mathcal{I}(\omega)\equiv\sum_{i, j}\langle \cos(2\omega a_{ij})\rangle&=\sum_{i, j}\langle e^{2i\omega a_{ij}}\rangle\,,\\
}
and the 4-body expectation value
\es{exp2}{
\mathcal{J}(\omega,w)\equiv&\sum_{i,j,k,l}\left[\langle \cos(2\omega a_{ij})\cos(2w a_{kl})\rangle-\langle \cos(2\omega a_{ij})\rangle\langle\cos(2w a_{kl})\rangle\right]\\
&=\sum_{i,j,k,l}\left[\langle e^{2i\omega a_{ij}}e^{2iw a_{kl}}\rangle-\langle e^{2i\omega a_{ij}}\rangle\langle e^{2iw a_{kl}}\rangle\right]\,,
}
in terms of which we can write \eqref{4m} as 
  \es{4mApp}{
- \partial_m^4F^\text{pert}\vert_{m=0}=&-12\zeta(3)+\int_0^\infty d\omega\frac{8\omega^3 \mathcal{I}(\omega)}{\sinh^2\omega}+\int_0^\infty d\omega\int_0^\infty dw\frac{12w\omega \mathcal{J}(\omega,w)}{\sinh^2w\sinh^2\omega}\,.
 }
The 2-body term $\mathcal{I}(\omega)$ also occurs in the matrix model computation of the LHS of \eqref{constraint1}, whose non-instanton part can be written as \cite{Chester:2019pvm}
 \es{2m2L}{
\frac{c}{8}\frac{\partial_m^2\partial_\tau\partial_{\bar\tau}F^\text{pert}}{\partial_\tau\partial_{\bar\tau}F}=-\frac{1}{8\lambda^2}\partial_{\lambda^{-1}}^2\int_0^\infty d\omega\frac{\omega\mathcal{I}(\omega)}{\sinh^2\omega}\,.
 }
This quantity was actually already computed in \cite{Chester:2019pvm} to all orders in $1/N^2$ in the 't Hooft limit for finite $\lambda$ (and to any order in $1/\lambda$) using topological recursion \cite{Eynard:2004mh,Eynard:2008we}, and for finite $N,\lambda$ (ignoring instantons) using orthogonal polynomials \cite{mehta1981}. We will apply these methods to the 4-body term $\mathcal{J}(\omega,w)$, and then combine with the known results for $\mathcal{I}(\omega)$ to compute \eqref{4mApp}.

\subsection{$1/N^2$ expansion from topological recursion}
\label{top}

Following \cite{Chester:2019pvm}, we will relate the expectation values in \eqref{exp1} and \eqref{exp2} to expectation values of product of resolvents.  Let us define the $n$-point correlator as
 \es{RnDef}{
  R^n(y_1, \dots, y_n) \equiv N^{n-2} \left\langle\sum_{i_1} \frac{1}{y_1 -a_{i_1} }\cdots \sum_{i_n} \frac{1}{y_n -a_{i_n} }\right\rangle \,,
 }
where the expectation value is taken in the Gaussian matrix model \eqref{ZSUN}.  We can then write the expectation values in \eqref{exp1} and \eqref{exp2} in terms of inverse Laplace transforms of resolvents.  Defining the inverse Laplace transform of a function $f$ by 
\es{L}{
& \widehat f(b_1,\dots, b_n)\equiv \frac{1} {(2\pi i)^n} \left[\prod_{i=1}^n\int_{\gamma_i-i\infty}^{\gamma_i+i\infty}dy_i e^{b_i y_i} \right]f(y_1,\dots,y_n)\,,
}
with $\gamma_{i}$ chosen so that the contour lies to the right of all singularities in the integrand, we then have
 \es{IJToResolvents}{
  {\cal I}(\omega) &= \widehat R^2(2i \omega, -2i \omega) \,, \\
   {\cal J}(\omega) &= \widehat R^4(2i \omega, - 2i \omega, 2i w, -2iw)
    - \widehat R^2( 2 i \omega, - 2i \omega) \widehat R^2(2 i w, -2iw) \,.
 }

To compute the $2$-point and $4$-point functions appearing in \eqref{IJToResolvents}, we first use the fact that these quantities are equal to the analogous quantities defined in the $U(N)$ matrix model
 \es{ZUN}{
  Z_{U(N)} = \int d^{N} a\, e^{-\frac{8 \pi^2 N }{\lambda} \sum_i a_i^2}  \prod_{i < j}a_{ij}^2 \,,
 }
which differs from \eqref{ZSUN} only in that it does not have the delta function factor in the integrand.  Indeed, the expectation value of an operator ${\cal O}(a_{ij})$ that is invariant under $a_i \to a_i + i \alpha$, for some constant $\alpha$, is the same in the $U(N)$ and $SU(N)$ matrix models, as can be easily shown by considering the expression for $e^{ -\frac{8 \pi^2 \alpha^2}{\lambda}}  Z_{U(N)} \langle {\cal O}(a_{ij}) \rangle_{U(N)} = e^{ -\frac{8 \pi^2 \alpha^2}{\lambda}} \int d^N a \, e^{-\frac{8 \pi^2 N }{\lambda} \sum_i a_i^2}  \prod_{i < j}a_{ij}^2 {\cal O}(a_{ij})$, sending $a_i \to a_i + i \alpha$, and integrating over real $\alpha$.  In the formalism involving resolvents, a similar computation shows that the inverse Laplace transforms of an $n$-point function of resolvents in the $SU(N)$ and $U(N)$ matrix models are related by
 \es{RRelation}{
   \widehat R_{SU(N)}^n (b_1, \dots, b_n) =  \widehat R_{U(N)}^n (b_1, \dots, b_n) e^{ \frac{\lambda}{32 \pi^2 N^2} \left( \sum_i b_i \right)^2} \,.
 }
Thus, as long as the arguments of $\widehat R^n$ sum to zero, as is the case in \eqref{IJToResolvents}, there is no difference between the $U(N)$ and $SU(N)$ theories, so we will drop the subscript $U(N)$ in what follows. 

 The correlators of resolvents obey various relations similar to Ward identities in QFT\@.  In particular, the change of variables $a_i \to a_i + \epsilon \delta a_i$ with $\delta a_i = 1/(z - a_i)$, with $\epsilon$ infinitesimal, leads at first order in $\epsilon$ to the relation
 \es{R2Relation}{
  R^2(z, z) = \frac{16 \pi^2}{\lambda }N^2 ( z R^1(z)  - 1)  \,.
 }
The more complicated change of variables corresponding to $\delta a_i = \frac{1}{z - a_i} \prod_{j=1}^p \sum_{i_j} \frac{1}{w_p - a_{i_j}}$ leads to 
 \es{RpRelation}{
  \frac{1}{N^2} R^{p+2}(z, z, w_1, \ldots w_p)
   &+ \sum_{j=1}^p \frac{\partial}{\partial w_j} \frac{R^p(w_1, \ldots w_p) - R^p (w_1, \ldots w_{j-1}, z, w_{j+1}, \ldots, w_p)}{w_j - z} 
    \\
  &=  \frac{16 \pi^2}{\lambda }
   \left( z R^{p+1} (z, w_1, \ldots, w_p) - N^2 R^p(w_1, \ldots, w_p) \right)  \,.
 }
Eqs.~\eqref{R2Relation}--\eqref{RpRelation}, combined with large $N$ factorization properties, lead to recursion relations that allows one to determine $R^p$ recursively in $p$ and in $1/N$.  It is customary to write down these recursion relations in terms of the connected correlators 
 \es{W}{
W^n(y_1,\dots, y_n)\equiv
 R^n(y_1, \dots, y_n)_\text{conn} =  N^{n-2}\left\langle\sum_{i_1} \frac{1}{y_1 -a_{i_1} }\cdots \sum_{i_n} \frac{1}{y_n -a_{i_n} }\right\rangle_\text{conn.}\,.
}
In terms of the inverse Laplace transforms of $W^n$, we have
 \es{expToW}{
 \mathcal{I}(\omega)=N^2 \widehat W^1(2i\omega)\; \widehat W^1(-2i\omega)+ \widehat W^2(2i\omega,-2i\omega)\,,\\
}
and
\es{expToW2}{
\mathcal{J}(\omega,w)=&N^2\mathcal{J}^0(\omega,w)+\mathcal{J}^1(\omega,w)+N^{-2}\mathcal{J}^2(\omega,w)\,,\\
}
where we define
\es{expToW3}{
\mathcal{J}^0(\omega,w) &\equiv \widehat W^1(2i\omega)\;\widehat W^1(2iw)\;\widehat W^2(-2i\omega,-2iw)\\
&{}+\widehat W^1(2i\omega)\;\widehat W^1(-2iw)\;\widehat W^2(-2i\omega,2iw)\\
&{}+\widehat W^1(-2i\omega)\;\widehat W^1(2iw)\;\widehat W^2(2i\omega,-2iw)\\
&{}+\widehat W^1(-2i\omega)\;\widehat W^1(-2iw)\;\widehat W^2(2i\omega,2iw)\,,\\
}
\es{expToW4}{
\mathcal{J}^1(\omega,w) &\equiv \widehat W^2(2i\omega,2iw)\;\widehat W^2(-2i\omega,-2iw)\\
&{}+\widehat W^2(2i\omega,-2iw)\;\widehat W^2(-2i\omega,2iw)\\
&{}+\widehat W^1(2i\omega)\;\widehat W^3(-2i\omega,-2iw,2iw)\\
&{}+\widehat W^1(-2i\omega)\;\widehat W^3(2i\omega,-2iw,2iw)\\
&{}+\widehat W^1(2iw)\;\widehat W^3(-2i\omega,-2iw,2i\omega)\\
&{}+\widehat W^1(-2iw)\;\widehat W^3(2i\omega,-2i\omega,2iw)\,,\\
}
\es{expToW5}{
\mathcal{J}^2(\omega,w) &\equiv \widehat W^4(2i\omega,-2i\omega,2iw,-2iw)\,.\\
}

The resolvents can then be expanded in $1/N^2$ as 
\es{W2}{
W^n(y_1,\dots, y_n)\equiv\sum_{m=0}^\infty\frac{1}{N^{2m}} W^n_m(y_1,\dots, y_n)\,,
}
and each genus-$m$ term $W^n_m$ can be computed for finite $\lambda$ using a recursion formula in $n,m$ \cite{Eynard:2004mh,Eynard:2008we} starting with the base case $W^1_0$, as described e.g. in \cite{Chester:2019pvm}. We use resolvents up to $n+m\leq 5$, which we give in an attached \texttt{Mathematica} file.\footnote{Some of these resolvents were already shown in Appendix B of \cite{Chester:2019pvm}.} We then take the inverse Laplace transform in \eqref{expToW3} to get the $1/N^2$ expansion at finite $\lambda$ for $\mathcal{J}(\omega,w)$ in terms of integrals over the Fourier variables $w,\omega$ from \eqref{KFourier}. For instance, at leading order in $1/N^2$ we need only consider the genus-zero resolvents in $\mathcal{J}^0(\omega,w)$, which give
  \es{tree221}{
  \mathcal{J}^0(\omega,w)\big\vert_{N^2}=&\frac{8 \pi  
   J_1(\frac{\sqrt{\lambda } \omega }{\pi })
   J_1(\frac{w \sqrt{\lambda }}{\pi })}{\sqrt{\lambda }
   (w^2-\omega^2 ) }\left[\textstyle\omega  J_0\left(\frac{\sqrt{\lambda } \omega }{\pi }\right)
   J_1\left(\frac{w \sqrt{\lambda }}{\pi }\right)-w
   J_1\left(\frac{\sqrt{\lambda } \omega }{\pi }\right)
   J_0\left(\frac{w \sqrt{\lambda }}{\pi }\right)\right]\,.
 }
 We can then plug this expression, along with the leading order term in $\mathcal{I}(\omega)$ as given in Appendix B of \cite{Chester:2019pvm}, into \eqref{4m} to get the leading order in $N^2$ result at finite $\lambda$:
  \es{4mLead}{
&- \partial_m^4F^\text{pert}\vert_{m=0}=N^2\Bigg[\int_0^\infty d\omega\frac{32\omega \pi ^2  J_1(\frac{ \sqrt{\lambda }\omega}{\pi
   }){}^2}{\lambda\sinh^2\omega }\\
   &+\int_0^\infty d\omega \int_0^\infty dw\frac{96w\omega \pi  
   J_1(\frac{\sqrt{\lambda } \omega }{\pi })
   J_1(\frac{w \sqrt{\lambda }}{\pi })}{\sinh^2w\sinh^2\omega\sqrt{\lambda }
   (w^2-\omega^2 ) }\left[\textstyle\omega  J_0\left(\frac{\sqrt{\lambda } \omega }{\pi }\right)
   J_1\left(\frac{w \sqrt{\lambda }}{\pi }\right)-w
   J_1\left(\frac{\sqrt{\lambda } \omega }{\pi }\right)
   J_0\left(\frac{w \sqrt{\lambda }}{\pi }\right)\right]\Bigg] \\
   &{}+O(N^0)\,.
 }
 In the attached \texttt{Mathematica} file, we give explicit formulae for $\mathcal{J}(\omega,w)$ to order $O(N^{-4})$, which similarly take the form of four Bessel functions, while the expressions for $\mathcal{I}(\omega)$ were already given in Appendix B of \cite{Chester:2019pvm} and consist of two Bessel functions.
 
  We would also like to take the large $\lambda$ expansion of these results, so that we can apply them to the strong coupling expansion of the integrated correlator. For the first term in \eqref{4mApp}, which depends on $\mathcal{I}(\omega)$, the large $\lambda$ expansion can be performed just as in \cite{Chester:2019pvm}, and yields
  \es{firstTerm}{
\int_0^\infty d\omega\frac{8\omega^3 \mathcal{I}(\omega)}{\sinh^2\omega}=&N^2\Big[\frac{16 \pi ^2}{\lambda
   }-\frac{32 \pi ^2}{\lambda ^{3/2}}+\frac{24 \pi ^2 \zeta (3)}{\lambda ^{5/2}}+O(\lambda^{-\frac72})\Big]\\
   &+\Big[\frac{4 \pi ^2
   \sqrt{\lambda }}{15}-\frac{13 \pi ^2}{16 \lambda ^{3/2}}-\frac{75 \pi ^2 \zeta (3)}{32 \lambda
   ^{5/2}}+O(\lambda^{-\frac72})\Big]\\
   &+N^{-2}\Big[-\frac{1}{504} \pi ^2 \lambda ^{3/2}+\frac{13 \pi ^2 \sqrt{\lambda }}{1920}+\frac{1533 \pi ^2}{8192
   \lambda ^{3/2}}+O(\lambda^{-\frac52})\Big]\\
      &+N^{-4}\Big[-\frac{\pi ^2 \lambda ^{5/2}}{38400}+\frac{25 \pi ^2 \lambda
   ^{3/2}}{129024}-\frac{511 \pi ^2 \sqrt{\lambda }}{327680}+O(\lambda^{-\frac32})\Big]+O(N^{-6})\,.
}
Note that none of these terms have the right powers of $\pi$ compared to the holographic correlator, so all of them must be cancelled against corresponding terms in the second term in \eqref{4mApp}. For these terms, which depend on $\mathcal{J}(\omega,w)$, every $W^n_m$ except $W^2_0$ factorizes in terms of their argument $y_i$, so the inverse Laplace transform can be easily taken and gives products of four Bessel functions of the form
\es{exampleProd}{
w^a\omega^bJ_{n_1}(\frac{\sqrt{\lambda}\omega}{\pi})J_{n_2}(\frac{\sqrt{\lambda}\omega}{\pi})J_{n_3}(\frac{\sqrt{\lambda }w}{\pi})J_{n_4}(\frac{\sqrt{\lambda }w}{\pi})
}
for various integers $a,b,n_i$. As described in Appendix D of \cite{Binder:2019jwn}, we can take the large $\lambda$ limit of these Bessel functions using the Mellin-Barnes form
\es{mbbessel}{
J_\mu(x)J_\nu(x) = \frac 1 {2\pi i}\int_{c-\infty i}^{c+\infty i}ds\frac{\Gamma(-s)\Gamma(2s+\mu+\nu+1)\left(\frac12 x\right)^{\mu+\nu+2s}}{\Gamma(s+\mu+1)\Gamma(s+\nu+1)\Gamma(s+\mu+\nu+1)} \,,
}
 where the integrals over $w,\omega$  can be separately done with factors $\csch^2{w},\csch^2{\omega}$ from \eqref{KFourier} by twice using the identity
\es{id}{
\int_0^\infty d\omega\ \frac{\omega^{a}}{\sinh^2\omega} = \frac{1}{2^{a-1}}\Gamma(a+1)\zeta(a) \,,
}
 and then the contours can be closed to the left to get an expansion in $1/\lambda$. For these factorizable terms, using the finite $\lambda$ expressions for $\mathcal{J}$ in the attached \texttt{Mathematica} file, we get
\es{secondTermFac}{
\int_0^\infty d\omega\int_0^\infty dw\frac{12w\omega \mathcal{J}^\text{fac}(\omega,w)}{\sinh^2w\sinh^2\omega}=&\Big[-\frac{13 \sqrt{\lambda
   }}{6}+\frac{55}{12}+\frac{3  }{8\sqrt{\lambda}}+O(\lambda^{-1})\Big]\\
   &+N^{-2}\Big[\frac{23 \lambda ^{3/2}}{5760}-\frac{19 \lambda
   }{384}+\frac{125 \sqrt{\lambda }}{3072}+O(\lambda^{0})\Big]\\
      &+N^{-4}\Big[\frac{\lambda ^{7/2}}{552960}+\frac{83 \lambda
   ^3}{20736000}+\frac{59 \lambda ^{5/2}}{3440640}+O(\lambda^{2})\Big]+O(N^{-6})\,,
}
where we have only showed the terms that we will use, and recall that there is no leading order factorizable term. 
 
 The only exception to factorizability is the genus-zero 2-body resolvent $W^2_0(y_1,y_2)$\footnote{Note that $W^2_0(y_1,y_2)$ also appears in $\mathcal{I}(\omega)$, but in this case there is only one Fourier variable $w$ so the integral was always of the form \eqref{id}.} which takes the form
 \es{W20}{
W^2_0(y_1,y_2)=\frac{\frac{4 \pi ^2 y_1 y_2}{\lambda }-1- \sqrt{\frac{4 \pi ^2
   y_1^2}{\lambda }-1} \sqrt{\frac{4 \pi ^2 y_2^2}{\lambda }-1}}{2
   \left(y_1-y_2\right){}^2 \sqrt{\frac{4 \pi ^2 y_1^2}{\lambda }-1}
   \sqrt{\frac{4 \pi ^2 y_2^2}{\lambda }-1}}\,,
}
and whose inverse Laplace transform for distinct imaginary arguments is
\es{W20L}{
 \widehat W^2(2i\omega,2iw)=-\frac{w\omega\sqrt{\lambda}}{2\pi(w+\omega)}[ J_{0}(\frac{\sqrt{\lambda}\omega}{\pi})J_{1}(\frac{\sqrt{\lambda}w}{\pi})+J_{0}(\frac{\sqrt{\lambda}w}{\pi})J_{1}(\frac{\sqrt{\lambda}\omega}{\pi}) ]\,.
}
This term shows up in both $\mathcal{J}^0(\omega,w)$ and $\mathcal{J}^1(\omega,w)$, and so appears at every order in the large $N^2$ expansion of $\mathcal{J}(\omega,w)$. We do not know how to take the large $\lambda$ expansions of such terms, since the $w,\omega$ dependence does not factorize due to the $(w+\omega)$ in the denominator of \eqref{W20L}. For these terms, we instead performed the large $\lambda$ expansion numerically by evaluating the $w,\omega$ integrals at many value of $\lambda$ at high precision and fitting a curve. Using the expressions in the \texttt{Mathematica} file, we get
\es{secondTermNFac}{
\int_0^\infty d\omega\int_0^\infty &dw\frac{12w\omega \mathcal{J}^\text{non-fac}(\omega,w)}{\sinh^2w\sinh^2\omega}=\\
&N^2\Big[6+\frac{96 \zeta (3)}{\lambda ^{3/2}}-\frac{288 \zeta (5)}{\lambda
   ^{5/2}}-\frac{144 \zeta (3)^2}{\lambda
   ^{3}}-\Big(\frac{16 \pi ^2}{\lambda
   }-\frac{32 \pi ^2}{\lambda ^{3/2}}+\frac{24 \pi ^2 \zeta (3)}{\lambda ^{5/2}}\Big)+O(\lambda^{-\frac72})\Big]\\
&\Big[\Big(\frac{25}{6}-\frac{4 \pi ^2}{15}\Big)\sqrt{\lambda}+O(\lambda^{0})\Big]+N^{-2}\Big[\Big(\frac{\pi ^2}{504}-\frac{119}{5760}\Big)\lambda^{\frac32}+O(\lambda)\Big]\\
      &+N^{-4}\Big[-\frac{\lambda ^{7/2}}{552960}-\frac{1781\lambda^3}{145152000}+O(\lambda^{\frac52})\Big]+O(N^{-6})\,.
}
We can now combine \eqref{firstTerm}, \eqref{secondTermFac}, and \eqref{secondTermNFac} to get the final result 
\es{Ffinal}{
-\cF_4(\tau, \bar \tau) &= N^2\Big[6+\frac{96 \zeta (3)}{\lambda ^{3/2}}-\frac{288 \zeta (5)}{\lambda
   ^{5/2}}-\frac{144 \zeta (3)^2}{\lambda
   ^{3}}+O(\lambda^{-\frac72})\Big]\\
   &\quad+\Big[2\sqrt{\lambda}+O(\lambda^{0})\Big]-N^{-2}\Big[\frac{\lambda^{\frac32}}{60}+O({\lambda})\Big]-N^{-4}\Big[\frac{\lambda^{3}}{120960}+O({\lambda^{\frac52}})\Big]+O(N^{-6})\,.
}
This $1/N^2$ expansion can then be converted to a $1/c=4/(N^2-1)$ expansion when comparing to the holographic correlator. Note that for the leading $N^2$ term we were able to do the numerical large $\lambda$ expansion to many orders, but for subleading terms in $1/N^2$ we were only able to accurately read off a couple orders in large $\lambda$ sufficient to get the terms shown here. 

\subsection{Finite $N$ from orthogonal polynomials}
\label{ortho2}

We can also compute $\mathcal{J}(\omega,w)$ at finite $N$ and $\lambda$ in terms of four finite sums using the method of orthogonal polynomials \cite{mehta1981}, as was already done for $\mathcal{I}(\omega)$ in \cite{Chester:2019pvm}. We start by writing $\mathcal{I}(\omega)$ and $\mathcal{J}(\omega,w)$ as
\es{exps}{
\mathcal{I}(\omega,w)&=N(N-1)\langle \cos(2\omega(a_1-a_2))\rangle+N\,,\\
\mathcal{J}(\omega,w)&=2N(N-1)\mathfrak{J}_{1212}+4N(N-1)(N-2) \mathfrak{J}_{1231} +N(N-1)(N-2)(N-3)\mathfrak{J}_{1234}\\
&\qquad-(\mathcal{I}(w)-N)(\mathcal{I}(\omega)-N)\,,\\
\mathfrak{J}_{ijkl}&\equiv\langle \cos(2\omega(a_i-a_j))\cos(2w(a_k-a_l))\rangle-\langle \cos(2\omega(a_i-a_j))\rangle\langle\cos(2w(a_k-a_l))\rangle\,.
}
We then introduce a family of polynomials $p_n(a)$ using the Hermite polynomials $H_n(x)$:
 \es{pn}{
 p_n(a)\equiv \left(\frac{\lambda}{32\pi^2 N}\right)^{\frac n2}H_n\left(\frac{4\pi \sqrt{N}a}{\sqrt{2\lambda}}\right)\,,
 }
 which are orthogonal with respect to the Gaussian measure 
 \es{ortho}{
 \int da\, p_m(a)p_n(a)e^{-\frac{8\pi^2N}{\lambda}a^2}=n!\left(\frac{\lambda}{16\pi^2 N}\right)^n\sqrt{\frac{\lambda}{8\pi N}}\delta_{mn}\equiv h_n\delta_{mn}\,.
 }
 As shown in \cite{Chester:2019pvm}, these orthogonal polynomials can be used to write the expectation value of an $n$-body operator as an $n$-dimensional integral:
 \es{nbodyExp}{
 \langle\cO_n(a)\rangle=\frac{1}{N!}\sum_{\sigma\in S_N}\sum_{\mu\in S_n}(-1)^{|\mu|}\int\left(\prod_{i=1}^n da_i\frac{p_{\sigma(i)-1}(a_i)p_{\mu(\sigma(i))-1}(a_i)}{h_{\sigma(i)-1}}e^{-\frac{8\pi^2N}{\lambda}a_i^2}\right)\cO_n(a)\,.
 }
 For the $n$-body operators in \eqref{exps} with $n=2,3,4$, we can perform the integrals in \eqref{nbodyExp} using the identity
 \es{identity}{
 \int_{-\infty}^\infty e^{-x^2+yx}H_m(x)H_n(x)=e^{\frac{y^2}{4}}2^m\sqrt{\pi}m!y^{n-m}L_m^{n-m}(-y^2/2)\,,
 }
 and the sums over permutations can be simplified to sums from 1 to $N$. For instance, $\mathcal{I}(\omega)$ was computed in this way in \cite{Chester:2019pvm} and yields 
  \es{firstExp2}{
\mathcal{I}(\omega)=&e^{\frac{-\omega^2\lambda}{4\pi^2 N}}\left[\left[L_{N-1}^1\left(\frac{\omega^2\lambda}{4\pi^2N}\right)\right]^2-\sum_{i,j=1}^N(-1)^{i-j}L_{i-1}^{j-i}\left(\frac{\omega^2\lambda}{4\pi^2N}\right)L_{j-1}^{i-j}\left(\frac{\omega^2\lambda}{4\pi^2N}\right)\right]+N\,,
 }
 where $L_a^b(x)$ are generalized Laguerre polynomials. The expression for $\mathcal{J}(\omega,w)$ similarly involves sums over $L_a^b(x)$, but takes a rather complicated form that we give in the attached Mathematica file. After plugging these terms into \eqref{4mApp} we can perform the sums and Fourier integrals for any finite $N$ and compare to the topological recursion results (as given in the attached Mathematica file). We find that they match even down to $N=2$ for a large range of $\lambda$,  which is a nontrivial check.

\section{Evaluating $I_2$ analytically}
\label{ImtauAn}

In this Appendix, we describe how to compute $I_2[\mathcal{G}(U,V)]$ using the Mellin transform defined in \eqref{MellinDef}. We begin by writing $I_2[\mathcal{G}(U,V)]$ in \eqref{constraint1} as an integral over $M(s,t)$:
\es{mellinInt}{
I_2[\cG(U,V)]&=  -\frac{2}{\pi}\int_{-i \infty}^{i \infty} \frac{ds\, dt}{(4 \pi i)^2} \left(
    \Gamma \left[2 - \frac s2 \right]^2 \Gamma \left[2 - \frac t2 \right]^2 \Gamma \left[ \frac{s+t}{2} \right]^2 \cM(s, t) \right.\\
&\left.  \qquad\qquad\qquad\qquad \times\int^\infty_0 dr \int_0^\pi d\theta\sin^2\theta (1 + r^2 - 2 r \cos \theta)^{\frac s2-2} r^{t - 1}  \right) \,.
    }
  The integrals over $r,\theta$ are standard one-loop integrals in four dimension, which can be done explicitly to get
  \es{mellinInt2}{
I_2[\cG(U,V)]&=-\frac18 \int_{-i \infty}^{i \infty} \frac{ds\, dt}{(2 \pi i)^2} \Gamma \left[2-\frac{s}{2}\right] \Gamma
   \left[\frac{s}{2}\right] \Gamma \left[2-\frac{t}{2}\right]
   \Gamma \left[\frac{t}{2}\right] \Gamma
   \left[-\frac{s}{2}-\frac{t}{2}+2\right] \Gamma
   \left[\frac{s+t}{2}\right]\cM(s, t) \,.
  }
  The integrals over $s,t$ can be done for polynomial $\cM(s, t) $ (or $\cM^\text{SG}(s,t)$) by twice applying the Barnes lemma:
\es{barnes}{
\int_{-i\infty}^{i\infty}\frac{ds}{2\pi i}\Gamma(a+s)\Gamma(b+s)\Gamma(c-s)\Gamma(d-s) = \frac{\Gamma(a+c)\Gamma(b+d)\Gamma(b+c)\Gamma(b+d)}{\Gamma(a+b+c+d)} \,,
}
which holds for contours for which the poles of each Gamma function lie either to the left or to the right of the contour. Applying this to \eqref{SintM} and \eqref{treePoly} we get the first line in \eqref{integrals}.

\section{Ward identities}
\label{WARDAPPENDIX}

\subsection{Ward identity for $\langle S S \bar P P \rangle$} 

As mentioned in the main text, conformal and R-symmetry invariance implies that the four-point function $\langle SS \bar P P \rangle$ takes the form given in \eqref{SSPP}.  The non-trivial information is encoded in the functions ${\cal R}_i(U, V)$, which are related by SUSY Ward identities to the functions ${\cal S}_i(U, V)$ defined in \eqref{FourPoint}.  The relations are given in Eq.~(B.6) of \cite{Binder:2019jwn}.

Since the $\langle SSSS \rangle$ correlator is split into a free part and a part depending on a single function ${\cal T}(U, V)$, one can also write ${\cal R}_i(U, V)$ reflecting this split, as we did in \eqref{PRSchem}.  The non-free part of ${\cal R}_i(U, V)$ is thus encoded in three differential operators ${\bf R}_i(U, V, \partial_U, \partial_V)$ that act on the function ${\cal T}(U, V)$ from the $\langle SSSS \rangle$ correlator.  From (B.6) of \cite{Binder:2019jwn}, we deduce that these differential operators are:
\es{SSPPward}{
 {\bf R}_1(U,V, \partial_U, \partial_V)
  &=\frac18\left[  2U(U-V-3)\partial_U V +U V (2 - U + 2 V)\partial_V^2 V \right.\\
&+U^2(U - 2 -2 V)\partial^2_U V - (4 V^2-4 + U[1 + U - 5 V] ) \partial_V V\\
&\left. -U(U - 2 -2 V)(U+V-1)\partial_V\partial_U V+8V\right]\,,\\
{\bf R}_2(U,V, \partial_U, \partial_V) &= \frac14\left[ (4V+2UV-2-2V^2 ) \partial_V V+U V (V-1)\partial_V^2 V\right.\\
&+U(1+U-V)\partial_UV+U(V-1)(U+V-1)\partial_V\partial_U V\\
&\left.+U^2(V-1)\partial^2_U V\right]\,,\\
{\bf R}_3(U,V, \partial_U, \partial_V) &= \frac18\left[ U(1+U-V ) \partial_V V+U^2 V \partial_V^2 V\right.\\
&\left.+U^2(U +V-1)\partial_V\partial_U V+U^3\partial^2_U V\right]\,.
}

Note that the differential operator ${\bf R}_1 + 3 {\bf R}_3$ takes the form
 \es{RSimp}{
  {\bf R}_1 + 3 {\bf R}_3 
   &= \frac{U^3}{4} 
    \biggl[ 
     2 + (-1 + U + 3 V) \partial_U + 4 V \partial_V 
      +  U V \partial_U^2 \\
       &{}+ V (-1 + U + V) \partial_U \partial_V 
     + V^2 \partial_V^2
    \biggr] \frac{1 + U + V}{U^2} \,.
 }

\subsection{Ward identity for $\langle P\overline PP\overline P\rangle$}
\label{wardPPPP}

Let us now move on to discussing $\langle P\bar PP\bar P\rangle$.  As mentioned in the main text, conformal symmetry and R-symmetry imply that this correlation function can be written as in \eqref{PPPP} in terms of three functions $\cP_i(U, V)$.  These functions must be related by Ward identities to the functions appearing in the $\langle SSSS \rangle$ correlator.  To derive these relations, we use the component field method of \cite{Dolan:2001tt,Binder:2018yvd,Binder:2019jwn}. This method was already discussed in a closely related context in \cite{Binder:2019jwn}, so we will only present an outline of the derivation here.

\begin{table}
\hspace{-.4in}
\begin{tabular}{c||c|c|c|c|c|c|c|c|c}
 & ${S}$ & $\chi\,,\overline\chi$ & ${P\,,\overline P}$ & $j$ & $F\,,\overline F$& $\Psi\,,\overline\Psi$ & $T$ & $\lambda\,,\overline\lambda$ & ${\Phi}\,,\overline\Phi$\\
 \hline 
  $\Delta$   & $2$ & $\frac52$& $3$& $3$ & $3$& $\frac72$& $4$&  $\frac72$ & $4$  \\
 \hline
Spin $[j,j']$ &$[0,0]$ & $[\frac12,0]\,,[0,\frac12]$& $[0,0]$& $[\frac12,\frac12]$ & $[1,0]\,,[0,1]$& $[1,\frac12]\,,[\frac12,1]$ & $[1,1]$ & $[\frac12,0]\,,[0,\frac12]$ & $[0,0]$ \\
 \hline
${SU}(4)_R$  & ${\bf20'}$ & ${\bf20},{\bf\overline{20}}$ & $\overline{\bf {10}}\,,{\bf 10}$ & ${\bf15}$ & ${\bf6}$& ${\bf4},{\bf \bar 4}$& ${\bf1}$& $\bar{\bf4},{\bf4}$ & ${\bf1}$\\
  \hline
  ${U}(1)_B$ & $0$ & $\frac12\,,-\frac12$ & $1\,,-1$ & $0$ & $1\,,-1$& ${\frac12}\,,-\frac12$& ${0}$& ${\frac32},-\frac32$& ${2},-2$\\
\end{tabular}
\caption{Operators in the $\cN = 4$ stress energy tensor multiplet and their scaling dimensions $\Delta$, spins $[j,j']$ of the Euclidean Lorentz group $SO(4)\cong SU(2)\times SU(2)$, irreps of the $R$-symmetry group $SU(4)_R$, and charges of the bonus symmetry group $U(1)_B$.}
\label{stressTable}
\end{table}

In general, we can derive these Ward identities by first determining the most general forms of the four-point functions that are consistent with conformal symmetry and R-symmetry, and then imposing invariance only under the Poincar\'e  supercharges.  For $\langle P\overline PP\overline P\rangle$, the relevant Ward identity takes the schematic form
\es{var1}{
0=\bar\delta\langle P\overline P P\overline\chi\rangle=\langle P\overline PP\overline P\rangle+\langle \partial\chi \overline PP\overline\chi\rangle+\langle P \overline P\partial\chi\overline\chi\rangle+\langle P\overline P P\overline F\rangle+\langle P\overline P P\overline\lambda\rangle
\,,
}
where $\bar\delta$ denotes the action of the supercharge, and the other operators in the stress tensor multiplet are summarized in Table \ref{stressTable}. This Ward identity will give $\langle P\overline PP\overline P\rangle$ in terms of powers and derivatives of $U,V$ of $\langle \chi \overline PP\overline\chi\rangle$, which must be related to other correlators in a chain that will eventually reach $\langle SSSS\rangle$. These variations are
\es{var2}{
&0=\bar\delta\langle SSS\chi\rangle=\langle  \overline\chi SS\chi\rangle+\langle S \overline \chi S\chi\rangle+\langle SS \overline \chi \chi\rangle+\langle SSSj\rangle+\langle SSS\partial S\rangle\,,\\
&0=\bar\delta\langle SSP\overline\chi\rangle=\langle  \overline\chi SP\overline\chi\rangle+\langle S \overline \chi P\overline\chi\rangle+\langle SS \partial \chi \overline\chi\rangle+\langle SSP\overline P\rangle+\langle SSP\overline F\rangle\,,\\
&0=\bar\delta\langle S\overline PP\chi\rangle=\langle \overline\chi\overline PP\chi\rangle+\langle S\overline \lambda P\chi\rangle+\langle S\overline P\partial\chi\chi\rangle+\langle S\overline PPj\rangle+\langle S\overline PP\partial S\rangle\,,
}
where the first two were already considered in \cite{Binder:2019jwn}. Finally, we write $\langle SSSS\rangle$ in terms of $\mathcal{T}(U,V)$ using \eqref{FourPoint} to obtain the split of the quantities ${\cal P}_i(U, V)$ into a free part and an interacting part dependent on ${\cal T}(U, V)$ as in \eqref{PRSchem}.  Following the procedure outlined above, we find that the differential operators ${\bf P}_i(U, V, \partial_U, \partial_V)$ appearing in \eqref{PRSchem} are
\small
\es{ppppward1}{
{\bf P}_1 &= \frac{1}{4} \big[U^4 V^2\partial_U^2 \partial_V^2+2 U^4 V^2\partial_U^3 \partial_V+U^4 V^2
  \partial_U^4 +U^4\partial_U^2+3 U^4 V\partial_U^2 \partial_V+3 U^4 V\partial_U^3\\
   &+2
   U^3 V^3\partial_U \partial_V^3+4 U^3 V^3\partial_U^2 \partial_V^2+2 U^3 V^3\partial_U^3 \partial_V+12 U^3 V^2
  \partial_U \partial_V^2+15 U^3 V^2\partial_U^2 \partial_V-2 U^3 V^2\partial_U^2 \partial_V^2\\
   &+3 U^3 V^2
  \partial_U^3-2 U^3 V^2\partial_U^3 \partial_V+2 U^3\partial_U +14 U^3 V
  \partial_U \partial_V+7 U^3 V\partial_U^2-2 U^3\partial_U^2-6 U^3 V\partial_U^2 \partial_V\\
   &-3
   U^3 V\partial_U^3+U^2 V^4\partial_V^4+2 U^2 V^4\partial_U \partial_V^3+U^2 V^4
  \partial_U^2 \partial_V^2+9 U^2 V^3\partial_V^3+9 U^2 V^3\partial_U \partial_V^2\\
   &-2 U^2 V^3
  \partial_U \partial_V^3-2 U^2 V^3\partial_U^2 \partial_V^2+19 U^2 V^2\partial_V^2+2 U^2 V^2
  \partial_U \partial_V-9 U^2 V^2\partial_U \partial_V^2-2 U^2 V^2\partial_U^2\\
   &-3 U^2 V^2
  \partial_U^2 \partial_V+U^2 V^2\partial_U^2 \partial_V^2+4 V \left(2 U^2-4 U V+3 U+2 V^2-5 V+3\right)
  \partial_V-4 U^2 V\partial_U \\
   &+U^2\partial_U -5 U^2 V\partial_U \partial_V+3
   U^2 V\partial_U^2+U^2\partial_U^2+3 U^2 V\partial_U^2 \partial_V-3 U V^4
  \partial_V^3-3 U V^4\partial_U \partial_V^2\\
   &+4 V^4\partial_V^2-17 U V^3\partial_V^2+3
   U V^3\partial_V^3-4 U V^3\partial_U \partial_V+6 U V^3\partial_U \partial_V^2-8 V^3
  \partial_V^2+15 U V^2\partial_V^2\\
   &+2 U V^2\partial_U +13 U V^2
  \partial_U \partial_V-3 U V^2\partial_U \partial_V^2+4 V^2\partial_V^2-3 U V\partial_U -3 U
  \partial_U -9 U V\partial_U \partial_V+4\big]\,,\\
   }
   \es{ppppward2}{
   {\bf P}_2&= \frac{1}{4} U \big[\left(2 U^2-4 U V+U+2 V^2-3 V+1\right)\partial_V+V \left(10
   U^2+U (3-5 V)+(V-1)^2\right)\partial_V^2\\
   &+U \big(U^3\partial_U^2 \partial_V+U^3 V
  \partial_U^2 \partial_V^2+U^3\partial_U^3+2 U^3 V\partial_U^3 \partial_V+U^3 V\partial_U^4 +2 U^2
   V^2\partial_U \partial_V^3\\
   &+4 U^2 V^2\partial_U^2 \partial_V^2+2 U^2 V^2\partial_U^3 \partial_V+4 U^2
  \partial_U \partial_V+8 U^2 V\partial_U \partial_V^2+3 U^2\partial_U^2+13 U^2 V\partial_U^2 \partial_V-2
   U^2\partial_U^2 \partial_V\\
   &-2 U^2 V\partial_U^2 \partial_V^2+5 U^2 V\partial_U^3-U^2
  \partial_U^3-2 U^2 V\partial_U^3 \partial_V+U V^3\partial_V^4+2 U V^3\partial_U \partial_V^3\\
   &+U
   V^3\partial_U^2 \partial_V^2-V^3\partial_U \partial_V^2+11 U V^2\partial_U \partial_V^2-2 U V^2
  \partial_U \partial_V^3+4 U V^2\partial_U^2 \partial_V-2 U V^2\partial_U^2 \partial_V^2-2 V^2\partial_U \partial_V\\
   & +V^2 (7 U-V+1)
  \partial_V^3+2 V^2\partial_U \partial_V^2+10 U V\partial_U \partial_V-3 U
  \partial_U \partial_V-7 U V\partial_U \partial_V^2+3 U V\partial_U^2\\
   &-U\partial_U^2-5 U V
  \partial_U^2 \partial_V+U\partial_U^2 \partial_V+U V\partial_U^2 \partial_V^2+3 V
  \partial_U \partial_V- \partial_U \partial_V -V\partial_U \partial_V^2\big)\big]\,,\\
   }
   \es{ppppward3}{
   {\bf P}_3&=
  \frac{1}{4} \big[\left(-2 U^3-(4 V+1) U^2-\left(-14 V^2+9 V+1\right) U-4 (V-1)^2 (2
   V-1)\right)\partial_V-U \big( U^4 \partial_U^2 \partial_V\\
   &+V\partial_U^2 \partial_V^2
   U^4+ U^4 \partial_U^3+2 V U^4 \partial_U^3 \partial_V+V U^4 \partial_U^4 +4
   U^3 \partial_U \partial_V +8 V U^3 \partial_U \partial_V^2\\
   &+2 V^2 U^3 \partial_U \partial_V^3+4 U^3 \partial_U^2+16 V
   U^3 \partial_U^2 \partial_V -3 U^3 \partial_U^2 \partial_V+5 V^2 U^3 \partial_U^2 \partial_V^2 -3 V
   U^3 \partial_U^2 \partial_V^2 \\
   &+8 V U^3 \partial_U^3 -2 U^3 \partial_U^3+4 V^2 U^3 \partial_U^3 \partial_V-4 V
  U^3 \partial_U^3 \partial_V +V^2  U^3 \partial_U^4-V  U^3 \partial_U^4\\
   &+2
   U^2 \partial_U +24 V U^2 \partial_U \partial_V -5 U^2 \partial_U \partial_V+23 V^2 U^2 \partial_U \partial_V^2-13 V
   U^2 \partial_U \partial_V^2+4 V^3 U^2 \partial_U \partial_V^3\\
   &-4 V^2 U^2 \partial_U \partial_V^3+10 V
   U^2 \partial_U^2 -3 U^2 \partial_U^2+19 V^2 U^2 \partial_U^2 \partial_V-18 V
   U^2 \partial_U^2 \partial_V+3 U^2 \partial_U^2 \partial_V\\
   &+5 V^3 U^2 \partial_U^2 \partial_V^2-8 V^2
  U^2  \partial_U^2 \partial_V^2 +3 V U^2 \partial_U^2 \partial_V^2+3 V^2 U^2 \partial_U^3-4 V
   U^2 \partial_U^3+ U^2 \partial_U^3 \\
   &+2 V^3 U^2 \partial_U^3 \partial_V-4 V^2
   U^2 \partial_U^3 \partial_V+2 V U^2 \partial_U^3 \partial_V+V^3 (U+V-1) U \partial_V^4-4 V
   U \partial_U \\
   &+ U \partial_U+2 V U \partial_U \partial_V -2 U \partial_U \partial_V+8 V^3 U \partial_U \partial_V^2-10
   V^2 U \partial_U \partial_V^2+2 V U \partial_U \partial_V^2+2 V^4 U \partial_U \partial_V^3\\
   &-4 V^3
   U \partial_U \partial_V^3+2 V^2 U \partial_U \partial_V^3-2 V^2 U \partial_U^2+3 V
   U \partial_U^2- U \partial_U^2 -V^2 U \partial_U^2 \partial_V+2 V U \partial_U^2 \partial_V\\
   &- U \partial_U^2 \partial_V+V^4
   U \partial_U^2 \partial_V^2-3 V^3 U \partial_U^2 \partial_V^2+3 V^2 U \partial_U^2 \partial_V^2-V
   U \partial_U^2 \partial_V^2 +2 V^2\partial_U -3 V
  \partial_U +\partial_U \\
   &+V^2 \left(7 U^2+(8 V-4) U-3 (V-1)^2\right)\partial_V^3-4 V^3\partial_U \partial_V+11 V^2\partial_U \partial_V-10 V
  \partial_U \partial_V+3\partial_U \partial_V\\
   &-3 V^4\partial_U \partial_V^2+9 V^3\partial_U \partial_V^2-9 V^2
  \partial_U \partial_V^2+3 V\partial_U \partial_V^2\big)+V ((1-14 V) U^2+(16 V^2-21 V+5)
   U-10 U^3\\
   &-4 (V-1)^3)\partial_V^2\big]\,.\\
}\normalsize
Note that the differential operators $2 {\bf P}_1 + 2 {\bf P}_2 + {\bf P_3}$ appearing in \eqref{d4mFT} can be simplified to 
 \es{PSimp}{
  2 {\bf P}_1 &+ 2 {\bf P}_2 + {\bf P_3}
   = \frac{4}{U^3} \biggl[
    U^3 V \partial_U^4+U \left(U^2+U (19
   V-2)+10 V^2-11 V+1\right)
   \partial_U^2 \partial_V \\
   &{}+\left(6 U^2+U (36 V-7)+6
   V^2-7 V+1\right) \partial_U \partial_V +U^2 (U+9
   V-1) \partial_U^3 \\
   &{}+2 U^2 V (U+V-1)
   \partial_U^3 \partial_V +U V \left(U^2+U (4
   V-2)+(V-1)^2\right)\partial_U^2 \partial_V^2 \\
   &{}+V
   \left(10 U^2+U (19 V-11)+(V-1)^2\right)
   \partial_U \partial_V^2 +U V^3
   \partial_V^4 \\
   &{}+2 U V^2 (U+V-1)
   \partial_U \partial_V^3 +V^2 (9 U+V-1)
   \partial_V^3 +2 U (3 U+9 V-2)
  \partial_U^2 \\
  &{}+(6 U+6 V-2)
   \partial_V +2 V (9 U+3 V-2)
   \partial_V^2 +(6 U+6 V-2)
   \partial_U
   \biggr] \frac{1 + U + V}{U^2} \,.
 }

\bibliographystyle{ssg}
\bibliography{N42loop}

\end{document}